\documentclass[a4paper,12pt]{article} 
\usepackage{graphicx} 
\usepackage{graphics} 
\usepackage{amsmath} 
\usepackage{amsfonts}  
\usepackage{amssymb} 

\usepackage[colorlinks=true, pdfstartview=FitV, linkcolor=blue,  citecolor=blue, urlcolor=blue]{hyperref}  

\newcommand{\dfr}{\widehat{d}} 
\newcommand{\Cs}{\mathbb{C}} 
\newcommand{\Ks}{\mathbb{K}}
\newcommand{\Lu}{\mathbb{L}} 
\newcommand{\Vsc}{\mathbb{V}} 
\newcommand{\Pl}{\mathbb{P}} 

\def\l{\left}
\def\r{\right}
\def\be{\begin{equation}}
\def\ee{\end{equation}}
\def\beq{\begin{equation}}
\def\eeq{\end{equation}}
\def\d{\partial}

\begin{document}

\title{Scale relativity and fractal space-time: theory and applications}
\author{L. Nottale\\{\small CNRS, LUTH, Paris Observatory and Paris-Diderot University} \\{\small 92190 Meudon, France}\\
{\small laurent.nottale@obspm.fr}}
\maketitle

\begin{abstract}
In the first part of this contribution, we review the development of the theory of scale relativity and its geometric framework constructed in terms of a fractal and nondifferentiable continuous space-time. This theory leads (i) to a generalization of possible physically relevant fractal laws, written as partial differential equation acting in the space of scales, and (ii) to a new geometric foundation of quantum mechanics and gauge field theories and their possible generalisations.

In the second part, we discuss some examples of application of the theory to various sciences, in particular in cases when the theoretical predictions have been validated by new or updated observational and experimental data. This includes predictions in physics and cosmology (value of the QCD coupling and of the cosmological constant), to astrophysics and gravitational structure formation (distances of extrasolar planets to their stars, of Kuiper belt objects, value of solar and solar-like star cycles), to sciences of life (log-periodic law for species punctuated evolution, human development and society evolution), to Earth sciences (log-periodic deceleration of the rate of California earthquakes and of Sichuan earthquake replicas, critical law for the arctic sea ice extent) and tentative applications to system biology. 
 \end{abstract}

\section{Introduction}

One of the main concern of the theory of scale relativity is about the foundation of quantum mechanics. As it is now well known, the principle of relativity (of motion) underlies the foundation of most of classical physics. Now, quantum mechanics, though it is harmoniously combined with special relativity in the framework of relativistic quantum mechanics and quantum field theories, seems, up to now, to be founded on different grounds. Actually, its present foundation is mainly axiomatic, i.e., it is based on postulates and rules which are not derived from any underlying more fundamental principle.  

The theory of scale relativity suggests an original solution to this fundamental problem. Namely, in its framework quantum mechanics may indeed be founded on the principle of relativity itself, provided this principle (applied up to now to position, orientation and motion) be extended to scales. One generalizes the definition of reference systems by including variables characterizing their scale, then one generalizes the possible transformations of these reference systems by adding, to the relative transformations already accounted for (translation, velocity and acceleration of the origin, rotation of the axes), the  transformations of these variables, namely, their relative dilations and contractions.  In the framework of such a newly generalized relativity theory, the laws of physics may be given a general form that transcends and includes both the classical and the quantum laws, allowing in particular to study in a renewed way the poorly understood nature of the classical to quantum transition.

A related important concern of the theory is the question of the geometry of space-time at all scales. In analogy with Einstein's construction of general relativity of motion, which is based on the generalization of flat space-times to curved Riemannian geometry, it is suggested, in the framework of scale relativity, that a new generalization of the description of space-time is now needed, toward a still continuous but now nondifferentiable and fractal geometry (i.e., explicitly dependent on the scale of observation or measurement). New mathematical and physical tools are therefore developed in order to implement such a generalized description, which goes far beyond the standard view of differentiable manifolds.
One writes the equations of motion in such a space-time as geodesics equations, under the constraint of the principle of relativity of all scales in nature. To this purpose, covariant derivatives are constructed that implement the various effects of the nondifferentiable and fractal geometry.

As a first theoretical step, the laws of scale transformation that describe the new dependence on resolutions of physical quantities are obtained as solutions of differential equations acting in the space of scales. This leads to several possible levels of description for these laws, from the simplest scale invariant laws to generalized laws with variable fractal dimensions, including log-periodic laws and log-Lorentz laws of ``special scale-relativity", in which the Planck scale is identified with a minimal, unreachable scale, invariant under scale transformations (in analogy with the special relativity of motion in which the velocity $c$ is invariant under motion transformations).

The second theoretical step amounts to describe the effects induced by the internal fractal structures of geodesics on motion in standard space (of positions and instants). Their main consequence is the transformation of classical dynamics into a generalized, quantum-like self-organized dynamics.
The theory allows one to define and derive from relativistic first principles both the mathematical and physical quantum tools (complex,  spinor,  bispinor, then multiplet wave functions) and the equations of which these wave functions are solutions: a Schrodinger-type equation (more generally a Pauli equation for spinors)  is derived as an integral of the geodesic equation in a fractal space, then Klein-Gordon  and Dirac equations in the case of a full fractal space-time. We then briefly recall that gauge fields and gauge charges  can also be constructed from a geometric re-interpretation of gauge transformations as scale transformations in fractal space-time. 

 In a second part of this review, we consider some applications of the theory to various sciences, particularly relevant to the questions of evolution and development. In the realm of physics and cosmology, we compare the various theoretical predictions obtained at the beginning of the 90's for the QCD coupling constant and for the cosmological constant to their present experimental and observational measurements.  In astrophysics, we discuss applications to the formation of gravitational structures over many scales, with a special emphasis on the formation of planetary systems and on the validations, on the new extrasolar planetary systems and on Solar System Kuiper belt bodies discovered since 15 years, of the theoretical predictions of scale relativity (made before their discovery). This is completed by a validation of the theoretical prediction obtained some years ago for the solar cycle of 11 yrs on other solar-like stars whose cycles are now measured. In the realm of life sciences, we discuss possible applications of this extended framework to the processes of morphogenesis and the emergence of prokaryotic and eukaryotic cellular structures, then to the study of species evolution, society evolution, embryogenesis and cell confinement. This is completed by applications in Earth sciences, in particular to a prediction of the Arctic ice rate of melting and to possible predictivity in earthquake statistical studies.

\section{Theory}

\subsection{Foundations of scale relativity theory}

The theory of  scale relativity is based on 
the giving up of the hypothesis of manifold differentiability. In this framework, the 
coordinate transformations are continuous but can be nondifferentiable. This implies several consequences \cite{Nottale1993}, leading to the following steps of construction of the theory:

(1) One can prove the following theorem \cite{Nottale1993,Nottale1996A,BenAdda2000,Cresson2001,Cresson2002}: a continuous and nondifferentiable curve 
is fractal in a general meaning, namely, its length is explicitly
dependent on a scale variable $\varepsilon$, i.e., ${\cal L} ={\cal L}(\varepsilon)$, and it diverges, ${\cal L} \rightarrow \infty$, when ${\varepsilon \rightarrow 0}$. This theorem can be readily extended to a continuous and nondifferentiable manifold, which is therefore fractal, not as an hypothesis, but as a consequence of the giving up of an hypothesis (that of differentiability).
 
 (2) The fractality of space-time \cite{Nottale1993,Ord1983,Nottale1984,Nottale1989} involves the scale 
dependence of the reference frames. One therefore adds to the 
usual variables defining the coordinate system, new 
variables $\varepsilon$ characterizing its `state of scale'. In particular, the coordinates themselves become functions of these scale variables, i.e., $X=X(\varepsilon)$.
   
 (3) The scale variables $\varepsilon$ can never be defined in an absolute way, but only in a relative way. Namely, only their ratio $\rho=\varepsilon'/\varepsilon$ does have a physical meaning. In experimental situations, these scales variables amount to the resolution of the measurement apparatus (it may be defined as standard errors, intervals, pixel size, etc...). In a theoretical analysis, they are the space and time differential elements themselves. This universal behavior leads to extend the principle of relativity in such a way that it applies also to the transformations (dilations and contractions) of these resolution variables \cite{Nottale1989,Nottale1992,Nottale1993}.
 
\subsection{Laws of scale transformation}\label{sl}

\subsubsection{Fractal coordinate and differential dilation operator}\label{diloperator}

Consider a variable length measured on a fractal curve, and, more generally, a non-differentiable (fractal) curvilinear coordinate 
${\cal L}(s,\varepsilon)$, that depends on some parameter $s$ which characterizes the position on the curve (it may be, e.g., a time coordinate), and on the resolution 
$\varepsilon$. Such a coordinate generalizes to nondifferentiable and fractal space-times the concept of curvilinear coordinates introduced for curved Riemannian space-times in Einstein's general relativity  \cite{Nottale1993}.  

Such a scale-dependent fractal length ${\cal L}(s,\varepsilon)$, remains finite and differentiable when $\varepsilon \neq 0$, namely, one can define a slope for any resolution $\varepsilon$, being aware that this slope is itself a scale-dependent fractal function. It is only at the limit $\varepsilon \rightarrow 0$ that the length is infinite and the slope undefined, i.e., that nondifferentiability manifests itself.

Therefore the laws of dependence of this length upon position and scale may be written in terms of a double differential calculus, i.e., it can be the solution of differential equations involving the derivatives of ${\cal L}$ with respect to both $s$ and $\varepsilon$. 

As a preliminary step, one needs to establish the relevant form of the scale variables and the way they intervene in scale differential equations. For this purpose, let us apply an infinitesimal dilation $d \rho$ to the resolution, which is therefore transformed as $\varepsilon \rightarrow 
\varepsilon '=\varepsilon(1+d\rho)$. The dependence on position is omitted at this stage  in order to simplify the notation. By applying this transformation to a fractal coordinate $\cal L$, one obtains, to first order in the differential element,
\beq
{\cal L}(\varepsilon ')={\cal L}(\varepsilon +\varepsilon \, d\rho)={\cal L}(\varepsilon)+\frac
{\partial {\cal L}(\varepsilon)} {\partial \varepsilon} \, \varepsilon \, d\rho=
(1+\tilde{D} \; d\rho) \, {\cal L}(\varepsilon),
\label{eq.1}
\eeq
where $\tilde{D}$ is, by definition, the dilation operator.

Since $d \varepsilon/ \varepsilon = d \ln \varepsilon$, the identification
 of the two last members of equation~(\ref{eq.1}) yields 
\beq
\tilde{D}=\varepsilon \; \frac{\partial} {\partial \varepsilon}=\frac{\partial}
 {\partial \ln \varepsilon} \; .
\label{eq.2}
\eeq
This form of the infinitesimal dilation operator shows that the natural variable for the resolution is $\ln \varepsilon$, and that the expected new differential equations will indeed involve quantities such as $\partial {\cal L}(s,\varepsilon)/\partial \ln \varepsilon$. This theoretical result agrees and explains the current knowledge according to which most measurement devices (of light, sound, etc..), including their physiological counterparts (eye, ear, etc..) respond according to the logarithm of the intensity (e.g., magnitudes, decibels, etc..).

\subsubsection{Self-similar fractals as solutions of a first order scale differential equation}\label{simple}

Let us start by writing the simplest possible differential equation of scale, then by solving it. We shall subsequently verify that the solutions obtained comply with the principle of relativity. As we shall see, this very simple approach already yields a fundamental result: it gives a foundation and an understanding from first principles for self-similar fractal laws, which have been shown by Mandelbrot and many others to be a general description of a large number of natural phenomena, in particular biological ones (see, e.g., \cite{Mandelbrot1982,Novak1998,Losa2002}, other volumes of these series and references therein). In addition, the obtained laws, which combine fractal and scale-independent behaviours, are the equivalent for scales of what inertial laws are for motion \cite{Mandelbrot1982}. Since they serve as a fundamental basis of description for all the subsequent theoretical constructions, we shall now describe their derivation in detail.

The simplest differential equation of explicit scale dependence which one can write is of first order and states that the variation of ${\cal L}$ under an infinitesimal scale transformation $d\ln\varepsilon$ depends only on ${\cal L}$ itself. Basing ourselves on the previous derivation of the form of the dilation operator, we thus write
\beq
\frac{\partial {\cal L}(s,\varepsilon)} {\partial \ln \varepsilon}=\beta({\cal L}).
\label{eq.3}
\eeq

The function $\beta$ is {\it a priori} unknown. However, still looking for the simplest form of such an equation, we expand 
$\beta(\cal L)$ in powers of $\cal L$, namely we write $\beta({\cal L})= a+b {\cal L} + ...$. Disregarding for the moment the $s$ dependence, we obtain, to the first order, the following linear equation, in which $a$ and $b$ are constants:
\beq
\frac{d {\cal L} }{d \ln \varepsilon}=a+b{\cal L} .
\label{eq.4}
\eeq
In order to find the solution of this equation, let us change the names of the constants as $\tau_F=-b$ and ${\cal L}_0=a/\tau_F$, so that $a+b{\cal L}=-\tau_F({\cal L}-{\cal L}_0)$. We obtain the equation
 \beq
 \frac{d {\cal L}}{{\cal L}-{\cal L}_0}=-\tau_F \; {d \ln \varepsilon}.
 \eeq
Its solution reads 
\beq
{\cal L}(\varepsilon) = {\cal L}_0\; \left \{1+ \left( \frac{\lambda}
{\varepsilon} \right) ^{\tau_F}\right\},
\label{eq.5}
\eeq
where $\lambda$ is an integration constant. This solution corresponds to a length measured on a fractal curve up to a given point. One can now generalize it to a variable length that also depends on the position characterized by the parameter $s$. One obtains
\beq
{\cal L}(s,\varepsilon) = {\cal L}_0(s)\; \left \{1+\zeta (s) \left( \frac{\lambda}
{\varepsilon} \right) ^{\tau_F}\right\},
\label{eq.5aa}
\eeq
in which, in the most general case, the exponent $\tau_F$ may itself be a variable depending on the position.

The same kind of result is obtained for the projections on a given axis of such a fractal length \cite{Nottale1993}. Let $X(s,\varepsilon)$ be one of these projections, it reads
\beq
X(s,\varepsilon) = x(s) \; \left \{1+\zeta_x (s) \left( \frac{\lambda}
{\varepsilon} \right) ^{\tau_F}\right\}.
\eeq
In this case $\zeta_x(s)$ becomes a highly fluctuating function which may be described by a stochastic variable.

 The important point here and for what follows is that the solution obtained is the sum of two terms, a classical-like, ``differentiable part" and a nondifferentiable ``fractal part", which is explicitly scale-dependent and tends to infinity when $\varepsilon \to 0$ \cite{Nottale1993,Celerier2004}. By differentiating these two parts in the above projection, we obtain the differential formulation of this essential result,
\beq
dX=dx+d \xi,
\eeq
where $dx$ is a classical differential element, while $d\xi$ is a differential element of fractional order. This relation plays a fundamental role in the subsequent developments of the theory.

Consider the case when $\tau_F$ is constant. In the asymptotic small scale regime, $\varepsilon \ll \lambda$, one obtains a power-law dependence on resolution that reads
\beq
{\cal L}(s,\varepsilon) = {\cal L}_0(s) \left( \frac{\lambda}{\varepsilon}\right)^{\tau_F}.
\label{eq.7}
\eeq
We recognize in this expression the standard form of a self-similar fractal behaviour with constant fractal dimension $D_F= 1+ \tau_F$, which have already been found to yield a fair description of many physical and biological systems \cite{Mandelbrot1982}. Here the topological dimension is $D_T=1$, since we deal with a length, but this can be easily generalized to surfaces ($D_T=2$), volumes ($D_T=3$), etc.., according to the general relation $D_F= D_T+ \tau_F$. The new feature here is that this result has been derived from a theoretical analysis based on first principles, instead of being postulated or deduced from a fit of observational data. 

It should be noted that in the above expressions, the resolution is a length interval, $\varepsilon= \delta X$ defined along the fractal curve (or one of its projected coordinate).  But one may also travel on the curve and measure its length on constant time intervals, then change the time scale. In this case the resolution $\varepsilon$ is a time interval, $\varepsilon= \delta t$. Since they are related by the fundamental relation
\beq
\delta X^{D_F} \sim \delta t,
\eeq
the fractal length depends on the time resolution as  
\beq
X(s,\delta t)= X_0(s) \times \l(\frac{T}{\delta t}\r)^{1-1/D_F}.
\eeq
An example of the use of such a relation is Feynman's result according to which the mean square value of the velocity of a quantum mechanical particle is proportional to $\delta t^{-1}$ \cite[p. 176]{Feynman1965}, which corresponds to a fractal dimension $D_F=2$, as later recovered by Abbott and Wise \cite{Abbott1981} by using a space resolution.

More generally, (in the usual case when  $\varepsilon=\delta X$), following Mandelbrot, the scale exponent $\tau_F = D_F - D_T$ can be defined as the slope of the $(\ln \varepsilon,\ln {\cal L})$ curve, namely
\beq
\tau_F = \frac{d \ln {\cal L}}{d \ln ({\lambda/ \varepsilon})} \; .
\label{eq.6}
\eeq
For a self-similar fractal such as that described by the fractal part of the above solution, this definition yields a constant value which is the exponent in Eq.~(\ref{eq.7}). However, one can anticipate on the following, and use this definition to compute an ``effective" or ``local" fractal dimension, now variable, from the complete solution that includes the differentiable and the nondifferentiable parts, and therefore a transition to effective scale independence. Differentiating the logarithm of Eq.~(\ref{eq.9}) yields an effective exponent given by
\beq
\tau_{\rm eff}=\frac{\tau_F}{1+(\varepsilon/\lambda)^{\tau_F}}.
\eeq
The effective fractal dimension $D_F=1+\tau_F$  therefore  jumps from the nonfractal value $D_F=D_T=1$ to its constant asymptotic value at the transition scale $\lambda$.

\subsubsection{Galilean relativity of scales}\label{grs}

We can now check that the fractal part of such a law is compatible with the principle of relativity extended to scale transformations of the resolutions (i.e., with the principle of scale relativity). It reads ${\cal L} = {\cal L}_0(\lambda/\varepsilon)^{\tau_F}$ (Eq.~\ref{eq.7}), and it is therefore a law involving two variables ($\ln {\cal L}$ and $\tau_F$) in function of one parameter ($\varepsilon$) which, according to the relativistic view, characterizes the state of scale of the system (its relativity is apparent in the fact that we need another scale $\lambda$ to define it by their ratio). More generally, all the following statements remain true for the complete scale law including the transition to scale-independence, by making the replacement of $\cal L$ by ${\cal L}-{\cal L}_0$. Note that, to be complete, we anticipate on what follows and consider {\it a priori} $\tau_F$ to be a variable, even if, in the simple law first considered here, it takes a constant value.

  Let us take the logarithm of  Eq.~(\ref{eq.7}). It yields $\ln ({\cal L}/{\cal L}_0)=\tau_F \, \ln (\lambda/\varepsilon)$. The two quantities $\ln {\cal L}$ and $\tau_F$ then transform, under a finite scale transformation $\varepsilon \rightarrow 
\varepsilon '= \rho \, \varepsilon$, as
 \beq
\ln \frac{{\cal L} (\varepsilon ')}{{\cal L}_0} =
 \ln \frac{{\cal L} (\varepsilon)}{{\cal L}_0} -
  \tau_F \ln \rho \; ,
\label{eq.8}
\eeq
and, to be complete,
\beq
\tau_F'= \tau_F.
\label{eq.9}
\eeq
These transformations have exactly the same mathematical structure as the Galilean group of motion transformation (applied here to scale rather than motion), which reads
\beq
x'=x-t\; v, \;\;\;  t'=t.
\eeq
This is confirmed by the dilation composition law,
 $\varepsilon \rightarrow \varepsilon ' \rightarrow \varepsilon ''$, which writes
\beq
\ln \frac{\varepsilon ''}{ \varepsilon}= \ln \frac{\varepsilon '}{ \varepsilon} +\ln \frac {\varepsilon ''}{ \varepsilon '}
\; ,
 \label{eq.10}
\eeq
and is therefore similar to the law of composition of velocities between three reference systems $K$, $K'$ and $K"$,
\beq
V^{\prime \prime}(K^{\prime \prime}/K)=V(K'/K)+V'(K^{\prime \prime}/K').
\eeq
Since the Galileo group of motion transformations is known to be the simplest group that implements the principle of relativity, the same is true for scale transformations. 

It is important to realize that this is more than a simple analogy:  the same physical problem is set in both cases, and is therefore solved under similar mathematical structures (since the logarithm transforms what would have been a multiplicative group into an additive group). Indeed, in both cases, it amounts to find the law of transformation of a position variable ($X$ for motion in a Cartesian system of coordinates, $\ln{\cal L}$ for scales in a fractal system of coordinates) under a change of the state of the coordinate system (change of velocity $V$ for motion and of resolution  $\ln \rho$ for scale), knowing that these state variables are defined only in a relative way. Namely, $V$ is the relative velocity between the reference systems $K$ and $K'$, and $\rho$ is the relative scale: note that $\varepsilon$ and $\varepsilon'$ have indeed disappeared in the transformation law, only their ratio remains. This remark founds the status of resolutions as (relative) ``scale velocities" and of the scale exponent $\tau_F$ as a ``scale time".

Recall finally that, since the Galilean group of motion is only a limiting case of the more general Lorentz group, a similar generalization is expected in the case of scale transformations, which we shall briefly consider in Sec.~\ref{ssr}.

\subsubsection{Breaking of scale invariance}\label{transition}

The standard self-similar fractal laws can be derived from the scale relativity approach. However, it is important to note that Eq.~(\ref{eq.9}) provides us with another fundamental result. Namely, it also contains a spontaneous breaking of the scale symmetry. Indeed, it is characterized by the existence of a transition from a fractal to a non-fractal behaviour at scales larger than some transition scale $\lambda$. The existence of such a breaking of scale invariance is also a fundamental feature of many natural systems, which remains, in most cases, misunderstood.  

The advantage of the way it is derived here is that it appears as a natural, spontaneous, but only effective symmetry breaking, since it does not affect the underlying scale symmetry. Indeed, the obtained solution is the sum of two terms, the scale-independent contribution (differentiable part), and the explicitly scale-dependent and divergent contribution (fractal part). At large scales the scaling part becomes dominated by the classical part, but it is still underlying even though it is hidden. There is therefore an apparent symmetry breaking, though the underlying scale symmetry actually remains unbroken.

The origin of this transition is, once again, to be found in relativity (namely, in the relativity of position and motion). Indeed, if one starts from a strictly scale-invariant law without any transition,  ${\cal L}={\cal L}_0 (\lambda/\varepsilon)^{\tau_F}$, then adds a  translation in standard position space (${\cal L} \rightarrow {\cal L} + {\cal L}_1$), one obtains
\beq
{\cal L}'={\cal L}_1+{\cal L}_0 \l(\frac{\lambda}{\varepsilon}\r)^{\tau_F}={\cal L}_1 \l\{1+\l(\frac{\lambda_1}{\varepsilon}\r)^{\tau_F}\r\}.
\eeq
Therefore one recovers the broken solution (that corresponds to the constant $a \neq 0$ in the initial scale differential equation). This solution is now asymptotically scale-dependent (in a scale-invariant way) only at small scales, and becomes independent of scale at large scales, beyond some relative transition $\lambda_1$ which is partly determined by the translation itself.

\subsubsection{Generalized scale laws}

\paragraph{Discrete scale invariance, complex dimension and log-periodic behaviour }\label{sec:logper}

Fluctuations with respect to pure scale invariance are potentially important, namely the log-periodic correction to power laws that is provided, e.g., by complex exponents or complex fractal dimensions. It has been shown that such a behaviour provides a very satisfactory and possibly predictive model of the time evolution of many critical systems, including earthquakes and market crashes (\cite{Sornette1998} and references therein). More recently, it has been applied to the analysis of major event chronology of the evolutionary tree of life \cite{Chaline1999,Nottale2000B,Nottale2002B}, of human development \cite{Cash2002} and of the main economic crisis of western and precolumbian civilizations \cite{Grou1987,Nottale2000B,Johansen2001,Grou2004}.

One can recover log-periodic corrections to self-similar power laws through the requirement of covariance (i.e., of form invariance of equations) applied to scale differential equations \cite{Nottale1997B}. Consider a scale-dependent function ${\cal L} (\varepsilon )$. In the applications to temporal evolution quoted above, the scale variable is identified with the time interval $| t -  t_c |$, where $t_c $ is the date of a crisis. Assume that ${\cal L}$  satisfies a first order differential equation, 
\begin{equation}
\label{20.}
\frac{d{\cal L}}{d\ln\varepsilon } -  \nu {\cal L}   =  0 ,
    \end{equation}
whose solution is a pure power law ${\cal L}({\varepsilon }) \propto  {\varepsilon }^\nu$ (cf Sect.~\ref{simple}). Now looking for corrections to this law, one remarks that simply incorporating a complex value of the exponent $\nu$ would lead to large log-periodic fluctuations rather than to a controllable correction to the power law. So let us assume that the right-hand side of Eq.~(\ref{20.}) actually differs from zero 
\begin{equation}
\label{21.}
\frac{d{\cal L} }{d\ln\varepsilon }  - \nu {\cal L} = \chi   .
 \end{equation}
 
We can now apply the scale covariance principle and require that the new function ${\chi  }$ be solution of an equation which keeps the same form as the initial equation 
\begin{equation}
\label{22.}
\frac{d\chi }{d\ln\varepsilon }  -  \nu' \chi  = 0  .
    \end{equation}
Setting ${\nu' }= {\nu} + {\eta }$, we find that ${\cal L}$ must be solution of a second-order equation
\begin{equation}
\label{23.}
\frac{d ^{2}{\cal L}}{(d\ln\varepsilon ) ^{2}}  -   (2 \nu + \eta )  \frac{d{\cal L} }{d\ln\varepsilon }  + \nu (\nu +\eta ) {\cal L} = 0  .
    \end{equation}
The solution reads ${\cal L}(\varepsilon )  = {a}  {\varepsilon }^{\nu}( 1 + {b} {\varepsilon }^{\eta  })$, and finally, the choice of an imaginary exponent ${\eta } = {i \omega }$  yields a solution whose real part includes a log-periodic correction:
\begin{equation}
\label{24.}
{\cal L}(\varepsilon )  = a  \, \varepsilon^\nu \, [ 1 + b \cos(\omega  \ln\varepsilon ) ].    
\end{equation}
As previously recalled in Sect.~\ref{transition}, adding a constant term (a translation) provides a transition to scale independence at large scales.

\paragraph{Lagrangian approach to scale laws}
\label{ss:lagrang}
In order to obtain physically relevant generalizations of the above simplest (scale-invariant) laws, a Lagrangian approach can be used in scale space, in analogy with its use to derive the laws of motion, leading to reverse the definition and meaning of the variables \cite{Nottale1997B}. 

This reversal is an analog to that achieved by Galileo concerning motion laws.  Indeed, from the Aristotle viewpoint, ``time is the measure of motion". In the same way, the fractal dimension, in its standard (Mandelbrot's) acception, is defined from the topological measure of the fractal object (length of a curve, area of a surface, etc..) and resolution, namely (see Eq.~\ref{eq.6})
\begin{equation}
t = \frac{x }{v}  \;\;\;  \leftrightarrow \;\;\;  \tau_F=D_F-D_{T} = \frac{d\ln{\cal L}}{d\ln(\lambda/\varepsilon)}. 
\end{equation}
In the case, mainly considered here, when $\cal L$ represents a length (i.e., more generally, a fractal coordinate), the topological dimension is $D_T=1$ so that $\tau_F=D_F-1$. With Galileo, time becomes a primary variable, and the velocity is deduced from space and time, which are therefore treated on the same footing, in terms of a space-time (even though the Galilean space-time remains degenerate because of the implicitly assumed infinite velocity of light). 

In analogy, the scale exponent $\tau_F=D_F-1$ becomes in this new representation a primary variable that plays, for scale laws, the same role as played by time in motion laws (it is called ``djinn" in some publications which therefore introduce a five-dimensional `space-time-djinn' combining the four fractal fluctuations and the scale time). 

Carrying on the analogy, in the same way as the velocity is the derivative of position with respect to time, $v=dx/dt$, we expect the derivative of $\ln {\cal L}$ with respect to scale time $\tau_F$ to be a ``scale velocity".  Consider as reference the self-similar case, that reads $\ln {\cal L}=\tau_F \ln(\lambda/\varepsilon)$. Derivating with respect to $\tau_F$, now considered as a variable, yields $ {d \ln {\cal L}}/{d \tau_F}=\ln({\lambda/\varepsilon})$, i.e., the logarithm of resolution.  By extension, one assumes that this scale velocity provides a new general definition of resolution even in more general situations, namely,
 \beq
{\Vsc} = \ln \left(  \frac{\lambda}{ \varepsilon}\right) = \frac{d \ln {\cal L}}{d \tau_F} 
\; .
\label{eq.11a}
\eeq
One can now introduce a scale Lagrange function ${\widetilde{L}}(\ln {\cal L}, {\Vsc}, \tau_F)$, from which a scale action is constructed
\beq
{\widetilde{S}} = \int_{\tau_1}^{\tau_2} {\widetilde{L}}(\ln {\cal L}, {\Vsc},
 \tau_F) \; d \tau_F.
\label{eq.12a}
\eeq
The application of the action principle yields a scale Euler-Lagrange equation that writes
\beq
\frac{d} {d \tau_F} \frac{\partial {\widetilde{L}}} {\partial{\Vsc}} =
 \frac{\partial {\widetilde{L}}} {\partial \ln {\cal L}} \; .
\label{eq.13a}
\eeq
One can now verify that, in the free case, i.e., in the absence of any ``scale force" (i.e., $\partial{\widetilde{L}} / \partial \ln {\cal L} = 0$), one recovers the standard fractal laws derived hereabove. Indeed, in this case the Euler-Lagrange equation becomes
\beq
 \partial{\widetilde{L}} / \partial {\Vsc} = {\rm const} \Rightarrow {\Vsc} = {\rm const}.
\label{eq.14a}
\eeq
which is the equivalent for scale of what inertia is for motion. Still in analogy with motion laws, the simplest possible form for the Lagrange function is a quadratic dependence on the scale velocity, (i.e., $ {\widetilde{L}}~\propto~{\Vsc}^2$).
The constancy of ${\Vsc} = \ln (\lambda/\varepsilon)$ means that it is independent of the scale time $\tau_F$. Equation (\ref{eq.11a}) can therefore be integrated to give the usual power law behaviour,
 ${\cal L} = {\cal L}_0 (\lambda/\varepsilon)^{\tau_F}$, as expected.
 
 But this reversed viewpoint has also several advantages which allow a full implementation of the principle of scale relativity: 

(i) The scale time $\tau_F$ is given the status of a fifth dimension and the logarithm of the resolution, ${\Vsc}=\ln(\lambda/\varepsilon)$, its status of scale velocity (see Eq.~\ref{eq.11a}). This is in accordance with its 
scale-relativistic definition, in which it characterizes the state of scale of the reference system, in the same way as the velocity 
$v = dx/dt$ characterizes its state of motion.

 (ii) This allows one to generalize the formalism to the case of four independent space-time resolutions, ${\Vsc}^{\mu} =
 \ln (\lambda^{\mu} / \varepsilon^{\mu}) = d \ln {\cal L}^{\mu}/ d \tau_F$.
 
(iii) Scale laws more general than the simplest self-similar ones can be
 derived from more general scale Lagrangians \cite{Nottale1997A,Nottale1997B} involving ``scale accelerations" ${\rm I}\!\Gamma= d^{2}\ln{\cal L} / d\tau_F^{2} = d \ln( \lambda /\varepsilon ) / d\tau_F $, as we shall see in what follows.

 Note however that there is also a shortcoming in this approach. Contrarily to the case of motion laws, in which time is always flowing toward the future (except possibly in elementary particle physics at very small time scales), the variation of the scale time may be non-monotonic, as exemplified by the previous case of log-periodicity. Therefore this Lagrangian approach is restricted to monotonous variations of the fractal dimension, or, more generally, to scale intervals on which it varies in a monotonous way.

\paragraph{Scale dynamics}

The previous discussion indicates that the scale invariant behaviour corresponds to freedom (i.e. scale force-free behaviour) in the framework of a scale physics. However, in the same way as there are forces in nature that imply departure from inertial, rectilinear uniform motion, we expect most natural fractal systems to also present distorsions in their scale behaviour with respect to pure scale invariance. This implies taking non-linearity in the scale space into account. Such distorsions may be, as a first step, attributed to the effect of a dynamics of scale (``scale dynamics"), i.e., of a ``scale field", but it must be clear from the very beginning of the description that they are of geometric nature (in analogy with the Newtonian interpretation of gravitation as the result of a force, which has later been understood from Einstein's general relativity theory as a manifestation of the curved geometry of space-time).

In this case the Lagrange scale-equation takes the form of Newton's equation of dynamics,
\begin{equation}
\label{15.}
F  =  \mu \;   \frac{d ^{2}\ln{\cal L}}{d\tau_F^{2}}   ,    
\end{equation}
where $\mu $ is a ``scale mass", which measures how the system resists to the scale force, and where ${\rm I}\!\Gamma = d^{2}\ln{\cal L} / d\tau_F^{2} = d \ln( \lambda /\varepsilon ) / d\tau_F $ is the scale acceleration. 

In this framework one can therefore attempt to define generic, scale-dynamical behaviours which could be common to very different systems, as corresponding to a given form of the scale force. 

\paragraph{Constant scale force}

A typical example is the case of a constant scale force. Setting $G=F/\mu$, the potential reads $\varphi  = G \ln{\cal L}$, in analogy with the potential of a constant force $f$ in space, which is $\varphi=-f x$, since the force is $-\d \varphi /\d x=f$. The scale differential equation  writes
\begin{equation}
\label{16.}
\frac{d ^{2}\ln{\cal L}}{d\tau_F   ^{2}}   = G .
    \end{equation}
It can be easily integrated. A first integration yields $d \ln {\cal L} / d \tau_F= G \tau_F +  {\Vsc} _{0}$, where ${\Vsc} _{0}$ is a constant. Then a second integration yields a parabolic solution (which is the equivalent for scale laws of parabolic motion in a constant field),
\begin{equation}
\label{17.}
{\Vsc}  =  {\Vsc} _{0} + G  \tau_F  \;\;\;    ;   \;\;\;  \ln{\cal L}  =  \ln{\cal L} _{0}  +  {\Vsc} _{0} \tau_F   +  \frac{1}{2} G \, \tau_F  ^{2 },    
\end{equation}
where $\Vsc=d \ln {\cal L} / d \tau_F= \ln ({\lambda }/{\varepsilon })$.

However the physical meaning of this result is not clear under this form. This is due to the fact that, while in the case of motion laws we search for the evolution of the system with time, in the case of scale laws we search for the dependence of the system on resolution, which is the directly measured observable. 
Since the reference scale $\lambda$ is arbitrary, the variables can be re-defined in such a way that $\Vsc_0=0$, i.e., $\lambda=\lambda_0$. Indeed, from Eq.~(\ref{17.}) one gets $\tau_F=(\Vsc-\Vsc_0)/G=[ \ln(\lambda/\varepsilon)-\ln (\lambda/\lambda_0)]/G=\ln (\lambda_0 / \varepsilon)/G$. Then one obtains
\begin{equation}
\label{18.}
\tau_F   =  \frac{1}{G}  \ln \left(\frac{\lambda_0 }{\varepsilon }\right), \;\;\;\;\; \ln \left(\frac{{\cal L}}{{\cal L} _{0}}\right)  =  \frac{1}{2G}  \ln ^{2} \left(\frac{\lambda_0 }{\varepsilon }\right).    
\end{equation}

The scale time $\tau_F$ becomes a linear function of resolution (the same being true, as a consequence, of the fractal dimension $D_F=1+\tau_F $), and the ($\ln{\cal L}, \ln\varepsilon $) relation is now parabolic instead of linear. Note that, as in previous cases, we have considered here only the small scale asymptotic behaviour, and that we can once again easily generalize this result by including a transition to scale-independence at large scale. This is simply achieved by replacing $\cal L$ by $({\cal L}-{\cal L}_0)$ in every equations.

There are several physical situations where, after careful examination of the data, the power-law models were clearly rejected since no constant slope could be defined in the ($\log{\cal L}, \log \varepsilon $) plane. In the several cases where a clear curvature appears in this plane, e.g., turbulence \cite{Dubrulle1997}, sandpiles \cite{Cafiero1995}, fractured surfaces in solid mechanics \cite{Carpinteri1996}, the physics could come under such a scale-dynamical description. In these cases it might be of interest to identify and study the scale force responsible for the scale distorsion (i.e., for the deviation from standard scaling).

\subsubsection{Special scale-relativity}\label{ssr}

Let us close this section about the derivation of scale laws of increasing complexity by coming back to the question of finding the general laws of scale transformations that meet the principle of scale relativity \cite{Nottale1992}. It has been shown in Sec.~\ref{grs} that the standard self-similar fractal laws come under a Galilean group of scale transformations. However, the Galilean relativity group is known, for motion laws, to be only a degenerate form of the Lorentz group. It has been proved that a similar result holds for scale laws \cite{Nottale1992,Nottale1993}.

The problem of finding the laws of linear transformation of fields in a scale transformation $\Vsc=\ln \rho$ ($ \varepsilon \rightarrow \varepsilon '$) amounts to finding four quantities, $ a(\Vsc), b(\Vsc), c(\Vsc)$, and $ d(\Vsc)$,  such that
\begin{equation}
\label{10a}
\ln  \frac{{\cal L}'}{{\cal L}  _{0}}  =   a(\Vsc)  \, \ln  \frac{{\cal L}}{{\cal L} _{0}}   +  b(\Vsc) \,  \tau_F     ,     
\end{equation}
$$\tau_F'   =  c(\Vsc) \,  \ln  \frac{{\cal L}}{{\cal L}  _{0}}   +  d(\Vsc) \, \tau_F .$$
Set in this way, it immediately appears that the current `scale-invariant' scale transformation law of the standard form of constant fractal dimension (Eq.~\ref{eq.8}), given by $ a = 1, b = \Vsc, c  = 0$  and $ d = 1$, corresponds to a Galilean group.

This is also clear from the law of composition of dilatations, $ \varepsilon \rightarrow \varepsilon '\rightarrow \varepsilon ''$, which has a simple additive form, 
\begin{equation}
\label{11}
\Vsc ''=  \Vsc + \Vsc' .    
\end{equation}
However the general solution to the `special relativity problem' (namely, find  $ a, b, c$ and $ d$ from the principle of relativity) is the Lorentz group \cite{LevyLeblond1976,Nottale1992}. This result has led to the suggestion of replacing the standard law of dilatation, $ \varepsilon  \rightarrow  \varepsilon ' = \varrho \times\varepsilon $ by a new Lorentzian relation, namely, for $ \varepsilon  < \lambda  _{0}$ and $ \varepsilon '< \lambda  _{0}$ 
\begin{equation}
\label{12}
\ln \frac{\varepsilon '}{\lambda  _{0}}  =  \frac{\ln(\varepsilon /\lambda  _{0}) +  \ln\varrho   }{1 + \ln\varrho  \, \ln(\varepsilon /\lambda  _{0}) / \ln ^{2}({\bf \Lambda}/\lambda_{0})}   .    
\end{equation}
This relation introduces a fundamental length scale $ {\bf \Lambda} $, which is naturally  identified, toward the small scales, with the Planck length (currently $ 1.6160(11) \times 10 ^{-35}$ m) \cite{Nottale1992},
\begin{equation}
\label{13}
{\bf \Lambda} = l_{\Pl}  =  (\hbar G/c ^{3}) ^{1/2} ,   
\end{equation}
and toward the large scales (for $ \varepsilon  > \lambda  _{0}$ and $ \varepsilon '> \lambda  _{0}$) with the scale of the cosmological constant, $\Lu=\Lambda^{-1/2}$ \cite[Chap. 7.1]{Nottale1993}.

As one can see from Eq.~(\ref{12}), if one starts from the scale $ \varepsilon  = {\bf \Lambda} $ and apply any dilatation or contraction $ \varrho $, one obtains again the scale $ \varepsilon '  = {\bf \Lambda} $, whatever the initial value of $ \lambda  _{0}$. In other words, $ {\bf \Lambda} $ can be interpreted as a limiting lower (or upper) length-scale, impassable, invariant under dilatations and contractions.

As concerns the length measured along a fractal coordinate which was previously scale-dependent as $ \ln({\cal L}/{\cal L} _{0}) = \tau_{0 } \, \ln(\lambda  _{0}/\varepsilon )$ for $ \varepsilon  < \lambda  _{0}$, it becomes in the new framework, in the simplified case when one starts from the reference scale ${\cal L}_0$
\begin{equation}
\label{14}
 \ln \frac{\cal L}{\cal L} _{0}  =   \frac{\tau_{0 } \, \ln(\lambda_{0}/\varepsilon )}{\sqrt{1 - \ln ^{2}(\lambda  _{0}/\varepsilon ) / \ln ^{2}(\lambda  _{0}/{\bf \Lambda} )}}    .    
\end{equation}
The main new feature of scale relativity respectively to the previous fractal or scale-invariant approaches is that the scale exponent $ \tau_F $  and the fractal dimension $ D_F = 1 + \tau_F $, which were previously constant ($ D_F = 2, \tau_F  = 1$ ), are now explicitly varying with scale, following the law (given once again in the simplified case when we start from the reference scale ${\cal L}_0$):
\begin{equation}
\label{15}
 \tau_F (\varepsilon )   =   \frac{\tau_0}{\sqrt{1 - \ln ^{2}(\lambda  _{0}/\varepsilon ) / \ln ^{2}(\lambda  _{0}/{\bf \Lambda} )}}    .    
\end{equation}
Under this form, the scale covariance is explicit, since one keeps a power law form for the length variation, ${\cal L}= {\cal L}_0 (\lambda/\varepsilon)^{\tau_F(\varepsilon)}$, but now in terms of a variable fractal dimension.

For a more complete development of special relativity, including its implications as regards new conservative quantities and applications in elementary particle physics and cosmology, see \cite{Nottale1992,Nottale1993,Nottale1996A,Nottale2007}.

The question of the nature of space-time geometry at Planck scale is a subject of intense work (see e.g. \cite{Amelino2002,Laperashvili2001} and references therein). This is a central question for practically all theoretical attempts, including noncommutative geometry \cite{Connes1994,Connes1998}, supersymmetry and superstrings theories \cite{Green1987,Polchinski1998}, for which the compactification scale is close to the Planck scale, and particularly for the theory of quantum gravity. 
Indeed, the development of loop quantum gravity by Rovelli and Smolin \cite{Rovelli1988} led to the conclusion that the Planck scale could be a quantized minimal scale in Nature, involving also a quantization of surfaces and volumes \cite{Rovelli1995}.

Over the last years, there has also been significant research effort aimed at the development of a `Doubly-Special-Relativity' \cite{Amelino2001} (see a review in \cite{Amelino2002}), according to which the laws of physics involve a fundamental velocity scale $c$ and a fundamental minimum length scale $L_p$, identified with the Planck length. 

The concept of a new relativity in which the Planck length-scale would become a minimum invariant length is exactly the founding idea of the special scale relativity theory \cite{Nottale1992}, which has been incorporated in other attempts of extended relativity theories \cite{Castro1997,Castro2000}. But, despite the similarity of aim and analysis, the main difference between the `Doubly-Special-Relativity' approach and the scale relativity one is that the question of defining an invariant length-scale is considered in the scale relativity/fractal space-time theory as coming under a relativity of scales. Therefore the new group to be constructed is a multiplicative group, that becomes additive only when working with the logarithms of scale ratios, which are definitely the physically relevant scale variables, as one can show by applying the Gell-Mann-Levy method to the construction of the dilation operator (see Sec.~\ref{diloperator}).

\subsection{Fractal space and quantum mechanics}
\label{fsqm}

The first step in the construction of a theory of the quantum space-time from fractal and nondifferentiable geometry, which has been described in the previous sections, has consisted of finding the laws of explicit scale dependence at a given ``point" or ``instant" (under their new fractal definition). 

The next step, which will now be considered, amount to write the equation of motion in such a fractal space(-time) in terms of a geodesic equation.  As we shall see, this equation takes, after integration the form of a Schr\"odinger equation (and of the Klein-Gordon and Dirac equations in the relativistic case). This result, first obtained in Ref.~\cite{Nottale1993}, has later been confirmed by many subsequent physical \cite{Nottale1996A,Nottale1997A,Dubois2000,Celerier2004} and mathematical works, in particular by Cresson and Ben Adda \cite{Cresson2001,Cresson2003,BenAdda2004,BenAdda2005} and Jumarie \cite{Jumarie2001,Jumarie2006,Jumarie2006A,Jumarie2007}, including attempts of generalizations using the tool of the fractional integro-differential calculus \cite{BenAdda2005,Cresson2007,Jumarie2007}. 

In what follows, we consider only the simplest case of fractal laws, namely, those characterized by a constant fractal dimension. The various generalized scale laws considered in the previous section lead to new possible generalizations of quantum mechanics \cite{Nottale1996A,Nottale2007}.

\subsubsection{Critical fractal dimension 2}
Moreover, we simplify again the description by considering only the case $D_F=2$. Indeed, the nondifferentiability and fractality of space implies that the paths are random walks of the Markovian type, which corresponds to such a fractal dimension. This choice is also justified by Feynman's result  \cite{Feynman1965}, according to which the typical paths of quantum particles (those which contribute mainly to the path integral) are nondifferentiable and of fractal dimension $D_F=2$ \cite{Abbott1981}. The  case $D_F\neq 2$, which yields generalizations to standard quantum mechanics has also been studied in detail (see \cite{Nottale1996A,Nottale2007} and references therein). This study shows that $D_F=2$ plays a critical role in the theory, since it suppresses the explicit scale dependence in the motion (Schr\"odinger) equation -- but this dependence remains hidden and reappears through, e.g., the Heisenberg relations and the explicit dependence of measurement results on the reolution of the measurement apparatus.

Let us start from the result of the previous section, according to which the solution of a first order scale differential equation reads for $D_F=2$, after differentiation and reintroduction of the indices,
 \begin{equation}
  \label{equ3}
dX^{\mu} = dx^{\mu} + d\xi^{\mu}= v^{\mu} ds + \zeta^{\mu} \sqrt{\lambda_c \, ds},
\end{equation}
where $\lambda_c$ is a length scale which must be introduced for dimensional reasons and which, as we shall see, generalizes the Compton length. The $\zeta^{\mu}$ are dimensionless highly fluctuating functions. 
Due to their highly erratic character, we can replace them by stochastic variables such that $<\!\!\zeta^{\mu}\!\!>=0$, $<\!\!(\zeta^0)^2\!\!>=-1$ and $<\!\!(\zeta^k)^2\!\!>=1$ ($k=$1 to 3). The mean is taken here on a purely mathematic probability law which can be fully general, since the final result odes not depend on its choice.
 
\subsubsection{Metric of a fractal space-time}
Now one can also write the fractal fluctuations in terms of the coordinate differentials,
$d \xi^{\mu}=\zeta^{\mu} \sqrt{\lambda^{\mu} \, dx^{\mu}}$.
The identification of this expression with that of Eq.~(\ref{equ3}) leads to recover the Einstein-de Broglie length and time scales,
\begin{equation}
\lambda_{x}=\frac{\lambda_c}{dx/ds}=\frac{\hbar}{p_x}, \;\;\;\tau=\frac{\lambda_c}{dt/ds}=\frac{\hbar}{E}.
\end{equation}

Let us now assume that the large scale (classical) behavior is given by Riemannian metric potentials ${g}_{\mu \nu}(x,y,z,t)$. The invariant proper time $dS$ along a geodesic writes, in terms of the complete differential elements $dX^{\mu}=dx^{\mu}+d \xi^{\mu}$, 
\begin{equation}
d{S}^2={g}_{\mu \nu} dX^{\mu} dX^{\nu}={g}_{\mu \nu} (dx^{\mu}+d \xi^{\mu}) (dx^{\nu}+d \xi^{\nu}).
\end{equation}
Now replacing the $d \xi$'s by their expression, one obtains a fractal metric \cite{Nottale1993,Nottale2001C}. Its two-dimensional and diagonal expression, neglecting the terms of zero mean (in order to simplify its writing) reads
\begin{equation}
d{S}^2={g}_{00}(x,t) \left( 1+ \zeta_{0}^2\; { \frac{\tau_F}{dt}}  \right)  c^2 dt^2-{g}_{11}(x,t) \left( 1+ \zeta_{1}^2\;{ \frac{\lambda_x}{dx}}  \right) dx^2.
\end{equation}

We therefore obtain generalized fractal metric potentials which are divergent and explicitly dependent on the coordinate differential elements \cite{Nottale1989,Nottale1993}. Another equivalent way to understand this metric consists in remarking that it is no longer only quadratic in the space-time differental elements, but that it also contains them in a linear way. 

As a consequence, the curvature is also explicitly scale-dependent and divergent when the scale intervals tend to zero. This property ensures the fundamentally non-Riemannian character of a fractal space-time, as well as the possibility to characterize it in an intrinsic way. Indeed, such a characterization, which is a necessary condition for defining a space in a genuine way, can be easily made by measuring the curvature at smaller and smaller scales. While the curvature vanishes by definition toward the small scales in Gauss-Riemann geometry, a fractal space can be characterized from the interior by the verification of the divergence toward small scales of curvature, and therefore of physical quantities like energy and momentum. 

Now the expression of this divergence is nothing but the Heisenberg relations themselves, which therefore acquire in this framework the status of a fundamental geometric test of the fractality of space-time \cite{Nottale1984,Nottale1989,Nottale1993}.
 
\subsubsection{Geodesics of a fractal space-time}

The next step in such a geometric approach consists in the identification of wave-particles with fractal space-time geodesics. 
Any measurement is interpreted as a selection 
of the geodesics bundle linked to the interaction with the measurement apparatus (that depends on its resolution) and/or to the information known about it (for example, the which-way-information in a two-slit experiment \cite{Nottale1996A}.

The three main consequences of nondifferentiability are: 

(i) The number of fractal geodesics is infinite. This leads to adopt a generalized 
statistical fluid-like description where the velocity $V^{\mu}(s)$ is 
replaced by a scale-dependent velocity field $V^{\mu}[X^{\mu}(s,ds),s,ds]$. 

(ii) There is a breaking of the reflexion invariance of the differential element $ds$. Indeed, in terms of fractal functions $f(s,ds)$, two derivatives are defined,
\begin{equation}
X'_+(s,ds) = \frac{X(s+ds,ds)-X(s,ds)}{ds}, \;\; 
X'_-(s,ds) = \frac{X(s,ds)-X(s-ds,ds)}{ds},
\end{equation}
which transform one in the other under the reflection ($ds \leftrightarrow -ds$), and which have a priori no reason to be equal. This leads to a fundamental two-valuedness of the velocity field.

(iii) The geodesics are themselves fractal curves of fractal dimension $D_F=2$ \cite{Feynman1965}. 

This means that one defines two divergent fractal
velocity fields, $V_+[x(s,ds),s,ds]=v_+[x(s),s] + w_+[x(s,ds),s,ds]$ and $V_-[x(s,ds),s,ds]=v_-[x(s),s]  + w_-[x(s,ds),s,ds]$, which can be decomposed in terms of differentiable parts $v_+$ and $v_-$, and of fractal parts $w_+$ and $w_-$. Note that, contrarily to other attempts such as Nelson's stochastic quantum mechanics which introduces forward and backward velocities \cite{Nelson1966} (and which has been later disproved \cite{Grabert1979,Wang1993}), the two velocities are here both forward, since they do not correspond to a reversal of the time coordinate, but of the time differential element now considered as an independent variable.

More generally, we define two differentiable parts of derivatives $d _{+}/ds$ and $d _{-}/ds$,
which, when they are applied to $x^\mu$, yield the differential parts of the velocity fields,
$ v^{\mu} _{+} ={{d _{+}} x^{\mu}/{ds}} $ and
$v^{\mu} _{-}={{d _{- }}x^{\mu} /{ds}}$.

\subsubsection{Covariant total derivative}

Let us first consider the non-relativistic case. It corresponds to a three-dimensional fractal space, without fractal time, in which the invariant $ds$ is therefore identified with the time differential element $dt$. One describes the elementary displacements $dX^k$, $k=1,2,3$, on the geodesics of a nondifferentiable fractal space in terms of the sum of two terms (omitting the indices for simplicity)
$dX_{\pm} = d_{\pm}x + d\xi_{\pm}$, where $dx$ represents the differentiable part and $d\xi$ the fractal (nondifferentiable) part, defined as
\begin{equation}
d_{\pm} x= v_{\pm} \; dt, \;\;\;
d\xi_{\pm}=\zeta_{\pm} \, \sqrt{2 \cal{D}}  \, dt^{1/2}.
\label{eq.20bis}
\end{equation}
Here $\zeta_{\pm}$ are stochastic dimensionless variables such that $<\!\!\zeta_{\pm}\!\!>=0$ and $<\!\!\zeta_{\pm}^2\!\!>=1$, and $\cal{D}$ is a parameter that generalizes, up to the fundamental constant $c/2$, the Compton scale (namely, ${\cal D}= \hbar/2m$ in the case of standard quantum mechanics). The two time derivatives are then combined in terms of a complex total time derivative operator \cite{Nottale1993},
\begin{equation}
\frac{\widehat{d}}{dt} = \frac{1}{2} \left( \frac{d_+}{dt} + \frac{d_-}{dt} \right) 
- \frac{i} {2} \left(\frac{d_+}{dt} - \frac{d_-}{dt}\right).
\label{eq.27}
\end{equation}
Applying this operator to the differentiable part of the position vector yields a complex velocity 
\begin{equation}
{\cal V} = \frac{\widehat{d}}{dt} x(t) = V -i U = \frac{v_+ + v_-}{2} - i 
\;\frac{v_+ - v_-}{2} \; .
\label{eq.28}
\end{equation} 
In order to find the expression for the complex time derivative operator, let us first calculate the derivative of a scalar function $f$. Since the fractal dimension is 2, one needs to go to second order of expansion. For one variable it reads
\beq
\frac{df}{ dt} = \frac{\partial f}{\partial t} + \frac{\partial f }{\partial X} \frac{dX}{dt} + 
\frac{1}{2} \frac{\partial ^2 f}{\partial X^2}\frac{ dX^2}
 {dt} .
\label{eq.30}
\eeq
The generalization of this writing to three dimensions is straighforward.

Let us now take the stochastic mean of this expression, i.e., we take the mean on the stochastic variables $\zeta_\pm$ which appear in the definition of the fractal fluctuation $d  \xi_\pm$. By definition, since $dX=dx+ d \xi$ and $<\!\! d\xi \!\!>=0$, we have $<\!\! dX\!\!>=dx$, so that the second term is reduced (in 3 dimensions) to $v. \nabla f$.  Now concerning the term $dX^2 /dt$, it is infinitesimal and therefore it would not be taken into account in the standard differentiable case. But in the nondifferentiable case considered here, the mean square fluctuation is non-vanishing and of order $dt$, namely, $<\!\!  d\xi^2 \!\!>=2 {\cal D} dt$, so that the last term of Eq.~(\ref{eq.30}) amounts in three dimensions to a Laplacian operator. One obtains, respectively for the (+) and (-) processes,   
\beq
\frac{d_\pm f} {dt} = \left(\frac{\partial } {\partial t} + v_\pm . \nabla
 \pm {\cal D} \Delta \right) f \; .
\label{eq.31}
\eeq
Finally, by combining these two derivatives in terms of the complex derivative of Eq.~(\ref{eq.27}), it reads \cite{Nottale1993}
\begin{equation}
\frac{\widehat{d}}{dt} =  \frac{\partial}{\partial t} + {\cal V}. \nabla -i {\cal  D} \Delta.
\end{equation} 
Under this form, this expression is not fully covariant \cite{Pissondes1999}, since it involves derivatives of the second order, so that its Leibniz rule is a linear combination of the first and second order Leibniz rules. By introducing the velocity operator \cite{Nottale2004A}
\begin{equation}
\widehat{\cal V}= {\cal V}-i {\cal D} \nabla,
\end{equation}
it may be given a fully covariant expression, 
\begin{equation}
\frac{\widehat{d}}{dt} =  \frac{\partial}{\partial t} +\widehat{\cal V}. \nabla,
\label{OK}
\end{equation}
namely, under this form it satisfies the first order Leibniz rule for partial derivatives. 

We shall now see that $\dfr/dt$ plays the role of a ``covariant derivative operator" (in analogy with the covariant derivative of general relativity), namely, one may write in its terms the equation of physics in a nondifferentiable space under a strongly covariant form identical to the differentiable case.

\subsubsection{Complex action and momentum}
The steps of construction of classical mechanics can now be followed, but in terms of complex and scale dependent quantities. One defines a Lagrange function that keeps its usual form, ${\cal L} (x, {\cal V}, t)$, but which is now complex, then a generalized complex action
\beq
{\cal S} = \int_{t_1}^{t_2} {\cal L} (x,{\cal V}, t) dt.
\label{eq.33}
\eeq
Generalized Euler-Lagrange equations that keep their standard form in terms of the new complex variables can be derived from this action \cite{Nottale1993,Celerier2004}, namely
\begin{equation}
\frac{\dfr}{dt} \, \frac{\partial L}{\partial {\cal V}} - \frac{\partial L}{\partial x}  =0 .
\end{equation}
From the homogeneity of space and Noether's theorem, one defines a generalized complex momentum given by the same form as in classical mechanics, namely,
\beq
{\cal P} = \frac{\partial {\cal L}}{\partial {\cal V}}.
\label{eq.35}
\eeq
If the action is now considered as a function of the upper limit of integration in Eq.~(\ref{eq.33}), the variation of the action from a trajectory to another nearby trajectory yields a generalization of another well-known relation of classical mechanics,
\begin{equation}
{\cal P} = \nabla {\cal S} .
\label{eq.36}
\end{equation}

\subsubsection{Motion equation}

Consider, as an example, the case of a single particle in an external scalar field of potential energy
$\phi$ (but the method can be applied to any situation described by a Lagrange function). The Lagrange function , $L=\frac{1}{2}mv^2 -\phi$,
 is generalized as ${\cal L} (x,{\cal V},t)=\frac{1}{2}m{\cal V}^2-\phi$.
The Euler-Lagrange equations then keep the form of Newton's fundamental equation
 of dynamics $F=m \,dv/dt$, namely,
\beq
m \, \frac{\dfr}{dt} \,{\cal V} =- \nabla \phi,
\label{eq.37}
\eeq
which is now written in terms of complex variables and complex operators.

 In the case when there is no external field ($\phi=0$), the covariance is explicit, since Eq.~(\ref{eq.37}) takes the free form of the equation of inertial motion, i.e., of a geodesic equation,
\beq
\frac{\dfr}{dt} \,{\cal V} = 0.
\label{eq.37bis}
\eeq
This is analog to Einstein's general relativity, where the equivalence principle leads to write the covariant equation of motion of a free particle under the form of an inertial motion (geodesic) equation $Du_{\mu}/ds=0$, in terms of the general-relativistic covariant derivative $D$, of the four-vector $u_{\mu}$ and of the proper time differential $ds$.

The covariance induced by the effects of the nondifferentiable geometry leads to an analogous transformation
 of the equation of motions, which, as we show below, become after integration the Schr\"odinger equation, which can therefore be considered as the integral of a geodesic equation in a fractal space.

In the one-particle case the complex momentum${\cal P}$ reads
\beq
{\cal P} = m {\cal V} ,
\label{eq.38}
\eeq
so that, from Eq.~(\ref{eq.36}), the complex velocity ${\cal V}$ appears as 
a gradient, namely the gradient of the complex action
\beq
{\cal V} = \nabla {\cal S}/ m.
\label{eq.39}
\eeq

\subsubsection{Wave function}

Up to now the various concepts and variables used were of a classical type (space, geodesics, velocity fields), even if they were generalized to the fractal and nondifferentiable, explicitly scale-dependent case whose essence is fundamentally not classical.

We shall now make essential changes of variable, that transform this apparently classical-like tool to quantum mechanical tools (without any hidden parameter or new degree of freedom). The complex wave function $\psi$ is introduced as simply another expression for the complex action ${\cal S}$, by making the transformation
\beq
\psi = e^{i{\cal S}/S_0}.
\label{eq.40}
\eeq
Note that, despite its apparent form, this expression involves a phase and a modulus since $\cal S$ is complex. The factor $S_0$ has the dimension of an action (i.e., an angular momentum) and must be introduced because $\cal S$ is dimensioned while the phase should be dimensionless. When this formalism is applied to standard quantum mechanics, 
${\cal S}_0$ is nothing but the fundamental constant $\hbar$. As a consequence, since
 \beq
{\cal S} = -i {\cal S}_0 \ln \psi,
\eeq
one finds that the function $\psi$ is related to the complex velocity appearing in Eq.~(\ref{eq.39}) as follows
\beq
{\cal V} = - i \, \frac{S_0}{m} \, \nabla \ln \psi.
\label{eq.41}
\eeq
This expression is the fondamental relation that connects the two description tools
while giving the meaning of the wave function in the new framework. Namely, it is defined here as a velocity potential for the velocity field of the infinite family of geodesics of the fractal space. Because of nondifferentiability, the set of geodesics that defines a `particle' in this framework is fundamentally non-local. It can easily be generalized to a multiple particle situation, in particular to entangled states, which are described by a single wave function $\psi$, from which the various velocity fields of the subsets of the geodesic bundle are derived as ${\cal V}_k = - i \, ({\cal S}_0/m_k) \, \nabla_k \ln \psi$, where $k$ is an index for each particle. The indistinguishability of identical particles naturally follows from the fact that the `particles' are identified with the geodesics themselves, i.e., with an infinite ensemble of purely geometric curves. In this description there is no longer any point-mass with `internal` properties which would follow a `trajectory', since the various properties of the particle  -- energy, momentum, mass, spin, charge (see next sections) -- can be derived from the geometric properties of the geodesic fluid itself.

\subsubsection{Correspondence principle}
Since we have ${\cal P} = -i {\cal S}_0 \nabla \ln  \psi= -i {\cal S}_0( \nabla  \psi)/\psi$,  we obtain the equality \cite{Nottale1993}
\beq
{\cal P} \psi= -i \hbar \nabla \psi
\eeq
in the standard quantum mechanical case ${\cal S}_0= \hbar$, which establishes a correspondence between the classical momentum $p$, which is the real part of the complex momentum in the classical limit, and the operator $ -i \hbar \nabla$.  

This result is generalizable to other variables, in particular to the Hamiltonian. Indeed, 
a strongly covariant form of the Hamiltonian can be obtained by using the fully covariant form Eq.~(\ref{OK}) of the covariant derivative operator. With this tool, the expression of the relation between the complex action and the complex Lagrange function reads
\beq
{\cal L}= \frac{\dfr \,{\cal S}}{dt}= \frac{\d {\cal S}}{\d t} + \widehat{\cal V}. \nabla {\cal S} \; .
\eeq
Since ${\cal P}= \nabla {\cal S}$ and ${\cal H}= -\d {\cal S}/\d t$, one obtains for the generalized complex Hamilton function the same form it has in classical mechanics, namely \cite{Nottale2007,Nottale2007A},
\beq
{\cal H}= \widehat{\cal V}. {\cal P} -{\cal L} \; .
\eeq
After expansion of the velocity operator, one obtains ${\cal H}={\cal V}. {\cal P} -i {\cal D} \nabla.{\cal P} -{\cal L}$, which includes an additional term \cite{Pissondes1999}, whose origin is now understood as an expression of nondifferentiability and strong covariance.

\subsubsection{Schr\"odinger equation and Compton relation}
The next step of the construction amounts to write the fundamental equation of dynamics Eq.~(\ref{eq.37}) in terms of the function $\psi$. It takes the form
\beq
i S_0\, \frac{\dfr}{dt}(\nabla \ln \psi) = \nabla \phi.
\label{eq.42}
\eeq
As we shall now see, this equation can be integrated in a general way under the form of a Schr\"odinger equation. Replacing $\dfr/dt$ and $\cal V$ by their expressions yields 
\beq
\nabla   \Phi  =   i S_0 \left[ \frac{\partial }{\partial t} \nabla
   \ln\psi   - i \left\{  \frac{S_0}{m} (\nabla   \ln\psi  . \nabla   )
(\nabla   \ln\psi ) + {\cal D} \Delta (\nabla   \ln\psi )\right\}\right] .
\label{eq.44}
\eeq
This equation may be simplified thanks to the identity \cite{Nottale1993},
\beq
\nabla\left(\frac{\Delta \psi}{\psi}\right)= 2 (\nabla \ln\psi . \nabla )
(\nabla \ln \psi )  + \Delta (\nabla \ln \psi).
\label{eq.50}
\eeq
We recognize, in the right-hand side of Eq.~(\ref{eq.50}), the two terms of
 Eq.~(\ref{eq.44}), which were respectively in factor of $S_0/{m}$
 and ${\cal D}$. This leads to definitely define the wave function as
\beq
\psi = e^{i{\cal S}/2m{\cal D}},
\label{eq.52}
\eeq
which means that the arbitrary parameter $S_0$ (which is identified with the constant $\hbar$ in standard QM) is now linked to the fractal fluctuation parameter by the relation
\beq
S_0=2 m {\cal D}.
\label{eq.51}
\eeq
This relation (which can actually be proved instead of simply being set as a simplifying choice, see \cite{Nottale2006A,Nottale2007A})  is actually a generalization of the Compton relation, since the geometric parameter ${\cal D}= <\!\! d \xi^2 \!\!>/2dt$ can be written in terms of a length scale as ${\cal D}= \lambda c/2$, so that, when $S_0= \hbar$, it becomes $\lambda=\hbar/mc$.
But a geometric meaning is now given to the Compton length (and therefore to the inertial mass of the particle) in the fractal space-time framework.

The fundamental equation of dynamics now reads
\beq
\nabla   \phi  =  2 i m {\cal D} \left[ \frac{\partial }{\partial t} \nabla
   \ln\psi   - i \left\{ 2 {\cal D} (\nabla   \ln\psi  . \nabla   )
(\nabla   \ln\psi ) + {\cal D} \Delta (\nabla   \ln\psi )\right\}\right] .
\eeq
Using the above remarkable identity and the fact that $\d/\d t$ and $ \nabla$ commute, it becomes
\beq
-\frac{\nabla \phi }{m}= -2 {\cal D} \nabla \left\{i \frac{\partial}
{\partial t} \ln \psi + {\cal D} \frac{\Delta \psi}{\psi}\right\} .
\label{eq.53}
\eeq
The full equation becomes a gradient,
 \beq
\nabla \l\{ \frac{\phi}{ m} -2 {\cal D} \nabla \left(   \frac{i \, {\partial \psi}/
{\partial t}+ {\cal D}{\Delta \psi}}{\psi} \r)  \r\}=0.
\eeq
and it can be easily integrated, to finally obtain a generalized Schr\"odinger equation \cite{Nottale1993}
\beq
{\cal D}^2 \Delta \psi + i {\cal D} \frac{\partial}{\partial t} \psi -
 \frac{\phi}{2m}\psi = 0,
\label{eq.54}
\eeq
up to an arbitrary phase factor which may be set to zero by a suitable choice of the $\psi$ phase. One recovers the standard Schr\"odinger equation of quantum mechanics for the particular case when ${\cal D}=\hbar/2m$.

\subsubsection{Von Neumann's and Born's postulates}
In the framework described here,  ``particles" are identified with the various geometric properties of fractal space(-time) geodesics. In such an interpretation, a measurement  (and more generally any knowledge about the system) amounts to a selection of the sub-set of the geodesics family in which are kept only the geodesics having the geometric properties corresponding to the measurement result. Therefore, just after the measurement, the system is in the state given by the measurement result, which is precisely the von Neumann postulate of quantum mechanics.

The Born postulate can also be inferred from the scale-relativity construction \cite{Celerier2004,Nottale2006A,Nottale2007A}. Indeed, 
 the probability for the particle to be found at a given position must be proportional to the density of the geodesics fluid at this point. The velocity and the density of the fluid are expected to be solutions of a Euler and continuity system of four equations, for four unknowns, $(\rho,V_x,V_y,V_z)$.

Now, by separating the real and imaginary parts of the Schr\"odinger equation, setting $\psi =\sqrt{P}\times e^{i \theta}$ and using a mixed representation ($P, V$), where $V=\{V_x,V_y,V_z\}$, one obtains precisely such a standard system of fluid dynamics equations, namely,
\begin{equation}
\label{AAB1}
 \left(\frac{\partial}{\partial t} + V \cdot \nabla \right) V  = -\nabla \left( {\phi}-2{\cal D}^2 \frac{\Delta \sqrt{P}}{\sqrt{P}}\right), \;\;\;\;\;
\frac{\partial P}{\partial t} + {\rm div}(P V) = 0.
\end{equation}
This allows one to univoquely identify $P=|\psi|^2$ with the probability density of the geodesics and therefore with the probability of presence of the `particle'. Moreover,
\begin{equation}
\label{Q1}
Q =-2{\cal D}^2 \frac{\Delta \sqrt{P}}{\sqrt{P}}
\end{equation}
can be interpreted as the new potential which is expected to emerge from the fractal geometry, in analogy with the identification of the gravitational field as a manifestation of the curved geometry in Einstein's general relativity. This result is supported by numerical simulations, in which the probability density is obtained directly from the distribution of geodesics without writing the Schr\"odinger equation \cite{Hermann1997,Nottale2007}. 

\subsubsection{Nondifferentiable wave function}
In more recent works, instead of taking only the differentiable part of the velocity field into account, one constructs the covariant derivative and the wave function in terms of the full velocity field, including its divergent nondifferentiable part of zero mean \cite{Nottale1999,Nottale2006A}. This still leads to the standard form of the Schr\"odinger equation. This means that, in the scale relativity framework, one expects the Schr\"odinger equation to have fractal and nondifferentiable solutions. This result agrees with a similar conclusion by Berry \cite{Berry1996} and Hall \cite{Hall2004}, but it is considered here as a direct manifestation of the nondifferentiability of space itself. The research of such a behavior in laboratory experiments is an interesting new challenge for quantum physics.

\subsection{Generalizations}

\subsubsection{Fractal space time and relativistic quantum mechanics}

All these results can be generalized to relativistic quantum mechanics, that corresponds in the scale relativity framework to a full fractal space-time. This yields, as a first step, the Klein-Gordon equation \cite{Nottale1994B,Nottale1996A,Celerier2004}.

Then the account of a new two-valuedness of the velocity allows one to suggest a geometric origin for the spin and to obtain the Dirac equation \cite{Celerier2004}. Indeed, the total derivative of a physical quantity also involves partial derivatives with respect to the space variables, $\partial / \partial x^{\mu}$. From the very definition of derivatives, the discrete symmetry under the reflection $dx^{\mu} \leftrightarrow -dx^{\mu}$ is also broken. Since, at this level of description, one should also account for parity as in the standard quantum theory, this leads to introduce a bi-quaternionic velocity field \cite{Celerier2004}, in terms of which Dirac bispinor wave function can be constructed.

We refer the interested reader to the detailed papers \cite{Nottale1996A,Celerier2004,Celerier2006}.

\subsubsection{Gauge fields as manifestations of fractal geometry}
 
The scale relativity principles has been also applied to the foundation of gauge theories, in the Abelian \cite{Nottale1994B,Nottale1996A} and non-Abelian \cite{Nottale2006,Nottale2007} cases.
 
This application is based on a general description of the internal fractal structures of the ``particle'' (identified with the geodesics of a nondifferentiable space-time) 
in terms of scale variables $\eta_{\alpha \beta}(x,y,z,t)=\varrho_{\alpha \beta}\,\varepsilon_{\alpha} \, \varepsilon_\beta$ whose true nature is tensorial, since it involves resolutions that may be different for the four space-time coordinates and may be correlated. This resolution tensor (similar to a covariance error matrix) generalizes the single resolution variable $\varepsilon $. Moreover, one considers here a more profound level of description in which the scale variables may now be function of the coordinates. Namely, the internal structures of the geodesics may vary from place to place and during the time evolution, in agreement with the non-absolute character of the scale space. 

This generalization amounts to construct a `general scale relativity' theory. The various ingredients of Yang-Mills theories (gauge covariant derivative, gauge invariance, charges, potentials, fields, etc...) can be  recovered in such a framework, but they are now founded from first principles and are found to be of geometric origin, namely, gauge fields are understood as manifestations of the fractality of space-time \cite{Nottale1994B,Nottale1996A,Nottale2006,Nottale2007}.

\subsubsection{Quantum mechanics in scale space}

One may go still one step further, and also give up the hypothesis of differentiability of the scale variables. Another generalization of the theory then amounts to use in scale space the method that has been built for dealing with nondifferentiability in space-time  \cite{Nottale2004B}.  This results in scale laws that take quantum-like forms instead of classical ones, and which may have several applications, as well in particle physics \cite{Nottale2004B} as in biology \cite{Nottale2007B}.

\section{Applications}

\subsection{Applications to physics and cosmology}

\subsubsection{Application of special scale relativity: value of QCD coupling}

In the special scale relativity framework, the new status of the Planck length-scale as a lowest unpassable scale must be universal. In particular, it applies also to the de Broglie and Compton relations themselves. They must therefore be generalized, since in their standard definition they may reach the zero length, which is forbidden in the new framework. 

A fundamental consequence of these new relations for high energy physics is that the mass-energy scale and the length-time scale are no longer inverse as in standard quantum field theories, but they are now related by the special scale-relativistic generalized Compton formula, that reads \cite{Nottale1992}
\begin{equation}
\ln \frac{m}{m_{0}}   = \frac{\ln (\lambda_{0}/\lambda )}{\sqrt{1 - {\ln ^{2}(\lambda_{0}/\lambda )}/{\ln ^{2}(\lambda_{0}/{l_\Pl} )}}}  .   
\label{newcompt}
\end{equation}

As a consequence of this new relation, one finds that the grand unification scale becomes the Planck energy scale \cite{Nottale1992, Nottale1993}. We have made the conjecture \cite{Nottale1992, Nottale1993} that the SU(3) inverse coupling reaches the critical value $4 \pi^{2}$ at this unification scale, i.e., at an energy $m_{\Pl}c^2/2 \pi$ in the special scale-relativistic modified standard model. 

 By running the coupling from the Planck to the $Z$ scale, this conjecture allows one to get a theoretical estimate for the value of the QCD coupling at $Z$ scale. Indeed its renormalization group equation yields a variation of $ \bar{\alpha }_{3}=\alpha_s$ with length scale given  to second order (for six quarks and $N_H$ Higgs doublets) by
\begin{eqnarray}
 \bar{\alpha }_{3}(r)  =  \bar{\alpha }_{3}(\lambda_{Z})  + \frac{7}{2\pi } \,\ln\frac{\lambda_{Z}}{r}\nonumber \\
   + \frac{11}{4\pi (40+N_{H})}\, \ln\left\{1 -  \frac{40+N_{H}}{20\pi } \,\alpha_{1}(\lambda_{Z})\ln\frac{\lambda_{Z}}{r}\right\}  \nonumber \\
-  \frac{27}{4\pi (20- N_{H})}\, \ln\left\{1 + \frac{20- N_{H}}{12\pi } \,\alpha_{2}(\lambda_{Z}) \ln\frac{\lambda_{Z}}{r}\right\} \nonumber\\
+ \frac{13}{14\pi }\, \ln\left\{1 + \frac{7}{2\pi }\, \alpha_{3}(\lambda_{Z})\ln\frac{\lambda_{Z}}{r}\right\}   .   
\end{eqnarray}

The variation with energy scale is obtained by making the transformation given by Eq.~(\ref{newcompt}). This led in 1992 to the expectation \cite{Nottale1992} $ \alpha_{3}(m_{Z}) = 0.1165\pm 0.0005$, that compared well with the experimental value at that time, $ \alpha_{3}(m_{Z}) = 0.112 \pm  0.010$, and was more precise. 

This calculation has been more recently reconsidered  \cite{Campagne2003,Nottale2007}, by using improved experimental values of the $\alpha_1$ and $\alpha_2$ couplings at $Z$ scale (which intervene at second order), and by a better account of the top quark contribution. Indeed, its mass was unknown at the time of our first attempt in 1992, so that the running from $Z$ scale to Planck scale was performed by assuming the contribution of six quarks on the whole scale range.

\begin{figure}[!ht]
\begin{center}
\includegraphics[width=14cm]{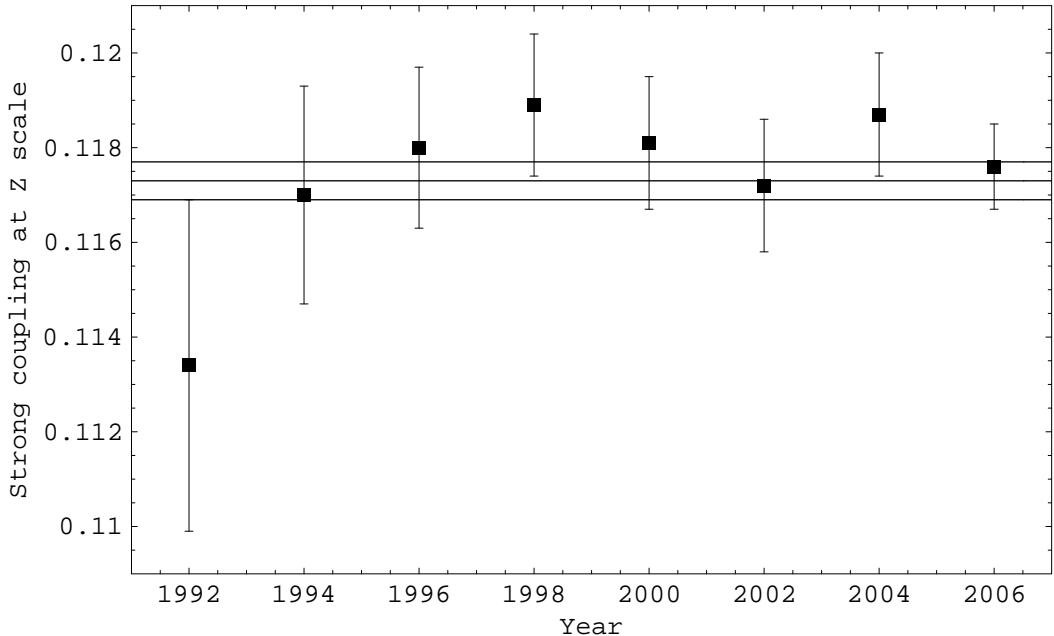}
\caption{\small Measured values of $\alpha_s(M_Z)$ from 1992 (date of the theoretical prediction) to 2006 \cite{PDG2006} compared with the expectation $\alpha_s(m_Z)=0.1173 \pm 0.0004$ made from assuming that the inverse running coupling reaches the value $4 \pi^2$ at Planck scale (see text).}
\label{alphasZ}
\end{center}
\end{figure}

However, the now known mass of the top quark,  $m_t=174.2 \pm 3.3$ GeV \cite{PDG2006} is larger than the $Z$ mass, so that only five quarks contribute to the running of the QCD coupling between $Z$ scale and top quark scale, then six quarks between top and Planck scale. Moreover, the possibility of a threshold effect at top scale cannot be excluded. This led to an improved  estimate :
\beq
\alpha_s(m_Z)=0.1173 \pm 0.0004,
\eeq
which agrees within uncertainties with our initial estimate $0.1165(5)$ \cite{Nottale1992}. This expectation is in very good agreement with the recent experimental average $\alpha_s(m_Z)=0.1176 \pm 0.0009$ \cite{PDG2006}, where the quoted uncertainty is the error on the average. We give in Fig.~\ref{alphasZ} the evolution of the measurement results of the strong coupling at $Z$ scale, which compare very well with the theoretrical expectation. 

\subsubsection{Value of the cosmological constant}
\label{cosmocst}
One of the most difficult open questions in present cosmology is the problem of the vacuum energy density and its manifestation as an effective cosmological constant. In the framework of the theory of scale relativity a new solution can be suggested to this problem, which also allows one to connect it to Dirac's large number hypothesis \cite[Chap. 7.1]{Nottale1993}, \cite{Nottale1996A}.

The first step toward a solution has consisted in considering the vacuum as fractal, (i.e., explicitly scale dependent). As a consequence, the Planck value of the vacuum energy density  is relevant only at the Planck scale, and becomes irrelevant at the cosmological scale.  One expects such a scale-dependent vacuum energy density to be solution of a scale differential equation that reads
\begin{equation}
\label{143.}
d\varrho /d \ln r   =   \Gamma (\varrho ) = a + b \varrho  + O(\varrho  ^{2})  ,     
\end{equation}
where $ \varrho $ has been normalized to its Planck value, so that it is always $ <1$, allowing a Taylor expansion of $ \Gamma (\varrho )$. This equation is solved as:
\begin{equation}
\label{144.}
\varrho  = \varrho_{c}  \left[  1 + \left(  \frac{r_0} {r } \right) ^{- b}  \right]  .     
\end{equation}
This solution is the sum of a fractal, power law behavior at small scales, that can be identified with the quantum scale-dependent contribution, and of a scale-independent term at large scale, that can be identified with the geometric cosmological constant observed at cosmological scales. The new ingredient here is a fractal/non-fractal transition about some scale $r_0$ that comes out as an integration constant, and which allows to connect the two contributions.

The second step toward a solution has been to realize that, when considering the various field contributions to the vacuum density, we may always chose $ <E> = 0$ (i.e., renormalize the energy density of the vacuum). But consider now the gravitational self-energy of vacuum fluctuations. It writes:
\begin{equation}
\label{145}
E_{g} =    \frac{G}{c ^{4}}  \frac{<E ^{2}>}{r}  .     
\end{equation}
The Heisenberg relations prevent from making $ <E ^{2}> = 0$, so that this gravitational self-energy {\it cannot} vanish. With $ <E ^{2}> ^{1/2}  =\hbar c /r$, we obtain the asymptotic high energy behavior:
\begin{equation}
\label{146}
\varrho_{g} = \varrho_{\Pl}   \left( \frac{l_\Pl }{r} \right) ^{6}   ,     
\end{equation}
where $ \varrho_{\Pl}$  is the Planck energy density and ${l_\Pl}$ the Planck length. From this equation one can make the identification $ - b  = 6$, so that one obtains $\varrho  = \varrho_{c}  \left[  1 + \left( {r_0} /{r } \right) ^6 \right]      $.

Therefore one of Dirac's large number relations is proved from this result \cite{Nottale1993}.
Indeed, introducing the characteristic length scale $ \Lu  = \Lambda  ^{- 1/2}$ of the cosmological constant $\Lambda$ (which is a curvature, i.e. the inverse of the square of a length), one obtains the relation:
\begin{equation}
\label{147}
\Ks =  \Lu /{l_\Pl }   = (r_{0}/{l_\Pl }) ^{3} = (m_{\Pl}/m_{0}) ^{3}  ,     
\end{equation}
where the transition scale $ r_{0}$ can be identified with the Compton length of a particle of mass $ m_{0}$. Then the power 3 in Dirac's large number relation is understood as coming from the power 6 of the gravitational self-energy of vacuum fluctuations and of the power 2 that relies the invariant  scale $ \Lu$ to the cosmological constant, following the relation $ \Lambda  = 1/ \Lu ^{2}$. The important point here is that in this new form of the Eddington-Dirac's relation, the cosmological length is no longer the time-varying $c/H_0$ (which led to theories of variation of constants), but the invariant cosmological length $\Lu$, which can therefore be connected to an invariant elementary particle scale without no longer any need for fundamental constant variation.

Now, a complete solution to the problem can be reached only provided the transition scale $r_{0}$ be identified. Our first suggestion \cite[Chap. 7.1]{Nottale1993} has been that this scale is given by the classical radius of the electron. 

Let us give an argument in favor of this conjecture coming from a description of the evolution of the primeval universe. The classical radius of the electron $r_e$ actually defines the $e^{+}e^{-}$ annihilation cross section and the $e^{-}e^{-}$ cross section $\sigma= \pi r_{\rm e}^2$ at energy $m_{\rm e} c^2$. This length corresponds to an energy $E_{\rm e}=\hbar c/r_{\rm e}=70.02$ MeV. This means that it yields the `size' of an electron viewed by another electron. Therefore, when two electrons are separated by a distance smaller than $r_{\rm e}$, they can no longer be considered as different, independent objects.

The consequence of this property for the primeval universe is that $r_{\rm e}$ should be a fundamental transition scale. When the Universe scale factor was so small that the interdistance between any couple of electrons was smaller than $r_{\rm e}$, there was no existing genuine separated electron. Then, when the cooling and expansion of the Universe separates the electron by distances larger than $r_{\rm e}$, the electrons that will later combine with the protons and form atoms appear for the first time as individual entities. Therefore the scale $r_{\rm e}$ and its corresponding energy 70 MeV defines a fundamental phase transition for the universe, which is the first appearance of electrons as we know them at large scales. Moreover, this is also the scale of maximal separation of quarks (in the pion), which means that the expansion, at the epoch this energy is reached, stops to apply to individual quarks and begins to apply to hadrons. This scale therefore becomes a reference static scale to which larger variable scales driven with the expansion can now be compared. Under this view, the cosmological constant would be a `fossil' of this phase transition, in similarity with the 3K microwave radiation being a fossil of the combination of electrons and nucleons into atoms. 

One obtains with the CODATA 2002 values of the fundamental constants a theoretical estimate
\begin{equation}
\Ks = (5.3000 \pm 0.0012) \times 10^{60},
\end{equation}
i.e. $\Cs_U=\ln \Ks=139.82281(22)$,
which corresponds to a cosmological constant (see \cite{Nottale1993} p. 305)
\begin{equation}
\Lambda = (1.3628 \pm 0.0004) \times 10^{-56}\; {\rm cm}^{-2}
\end{equation}
i.e., a scaled cosmological constant
\begin{equation}
\Omega_{\Lambda}= (0.38874 \pm 0.00012) \, h^{-2}.
\end{equation}
Finally the corresponding invariant cosmic length scale is theoretically predicted to be
\begin{equation}
\Lu=(2.77608\pm 0.00042) \;{\rm Gpc},
\label{petitlu}
\end{equation}
i.e., $\Lu=(8.5661\pm 0.0013)\times 10^{25}$ m. 

Let us compare these values with the most recent determinations of the cosmological constant (sometimes now termed, in a somewhat misleading way,  `dark energy'). The  WMAP three year analysis of 2006 \cite{Spergel2006} has given $h=0.73 \pm 0.03$ and $\Omega_{\Lambda}({\rm obs})=0.72 \pm 0.03$. 
These results, combined with the recent Sloan (SDSS) data \cite{Tegmark2006}, yield, assuming $\Omega_{\rm tot}=1$ (as supported by its WMAP determination, $\Omega_{\rm tot}=1.003 \pm 0.010$)
\beq
\Omega_\Lambda({\rm obs})=\frac{\Lambda c^2}{3 H_{0}^2}=0.761 \pm 0.017, \;\;\; h=0.730 \pm 0.019.
\eeq
Note that these recent results have also reinforced the cosmological constant interpretation of the `dark energy' with a measurement of the coefficient of the equation of state $w=-0.941 \pm 0.094$ \cite{Tegmark2006}, which encloses the value $w=-1$ expected for a cosmological constant.

\begin{figure}[!ht]
\begin{center}
\includegraphics[width=14cm]{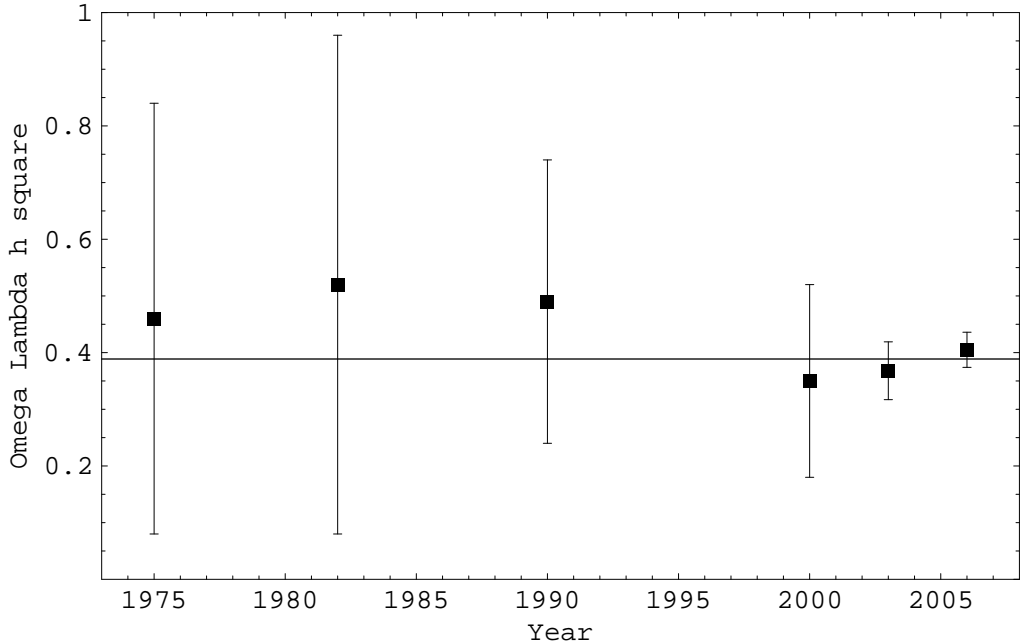}
\caption{\small  Evolution of the measured values of the dimensionless cosmological constant $\Omega_{\Lambda}h^2= {\Lambda c^2}/{3 H_{100}^2}$, from 1975 to 2006, compared to the theoretical expectation  $\Lambda=  ( m_{\rm e}/ \alpha \,m_\Pl  )^6 \;(1/l_\Pl)^2$ \cite{Nottale1993} that gives numerically $ \Omega_{\Lambda} h^{2}{\rm (pred)}= 0.38874 \pm 0.00012$.}
\label{OmegaYear}
\end{center}
\end{figure}

With these values one finds a still improved cosmological constant
\beq
\Omega_{\Lambda} h^{2}{\rm (obs)} =0.406 \pm 0.030,
\eeq
which corresponds to a cosmic scale
\beq
\Lu{\rm (obs)} =(2.72 \pm 0.10) \;{\rm Gpc},\;\;\; {\rm i.e.,}\; \Ks{\rm (obs)}= (5.19 \pm 0.19) \times 10^{60},
\eeq
in excellent agreement with the predicted values $\Lu{\rm (pred)}=2.7761(4)$ Gpc, and $\Ks= 5.300(1) \times 10^{60}$.

The evolution of these experimental determinations \cite{Nottale2007} is shown in Fig.~\ref{OmegaYear} where they are compared with the theoretical expectation 
\beq
\Omega_{\Lambda} h^{2}{\rm (pred)}= 0.38874 \pm 0.00012.
\eeq
The convergence of the observational values toward the theoretical estimate, despite an improvement of the precision by a factor of more than 20, is striking. The 2008 value from the Five-Year WMAP results is $\Omega_{\Lambda} h^{2}{\rm (obs)}= 0.384 \pm 0.043$ \cite{Hinshaw2008} and is once again in very good agreement with the theoretical expectation made 16 years ago \cite{Nottale1993}, before the first genuine measurements in 1998.

\subsection{Applications to astrophysics}
\subsubsection{Gravitational Schr\"odinger equation}

Let us first briefly recall the basics of the scale-relativistic theoretical approach. It has been reviewed in Sec.~\ref{fsqm} in the context of the foundation of microphysics quantum mechanics. We shall now see that some of its ingredients, leading in particular to obtain a generalized Schr\"odinger form for the equation of motion, also applies to gravitational structure formation. 

Under three general conditions, namely,  \{(i) infinity of geodesics (which leads to introduce a non-deterministic velocity field), (ii) fractal dimension $D_F=2$ of each geodesic, on which the elementary displacements are described in terms of the sum $dX=dx+d \xi$ of a classical, differentiable part $dx$ and of a fractal, non-differentiable fluctuation $d \xi$, (iii) two-valuedness of the velocity field, which is a consequence of  time irreversibility at the infinitesimal level issued from non-differentiability, one can construct a complex covariant derivative that reads
\begin{equation}
\frac{\dfr}{dt} = \frac{\partial}{\partial t} + {\cal V}. \nabla - i {\cal D} \Delta \; ,
\label{eq.32}
\end{equation}
where $\cal D$ is a parameter that characterizes the fractal fluctuation, which is such that $<d \xi^2>=2 {\cal D} dt$, and where the classical part of the velocity field, $\cal V$ is complex as a consequence of condition (iii) (see \cite{Celerier2004,Nottale2007A} for  more complete demonstrations).

Then this covariant derivative, that describes the non-differentiable and fractal geometry of space-time, can be combined with the covariant derivative of general relativity, that describes the curved geometry. We shall briefly consider in what follows only the Newtonian limit. In this case the equation of geodesics keeps the form of Newton's fundamental equation of dynamics in a gravitational field,

\begin{equation}
\frac{\widehat{D} {\cal V}}{dt}=\frac{\dfr{\cal V}}{dt} +\nabla \left(\frac{\phi}{m} \right) =0,
\end{equation}
where $\phi$ is the Newtonian potential energy.
Introducing the action $S$, which is now complex, and making the change of variable $\psi=e^{iS/2m{\cal D}}$, this equation can be integrated under the form of a generalized Schr\"odinger equation \cite{Nottale1993}:

\begin{equation}
\label{AB}
{\cal D}^2 \Delta \psi +  i{\cal D} \frac{\partial}{\partial t} \psi -\frac{\phi}{2m} \, \psi=0.
\end{equation}

Since the imaginary part of this equation is the equation of continuity (Sec.~3), and basing ourselves on our description of the motion in terms of an infinite family of geodesics, $P =| \psi|^2$ naturally gives the probability density of the particle position \cite{Celerier2004,Nottale2007A}. 

Even though it takes this Schr\"odinger-like form, equation (\ref{AB}) is still in essence an equation of gravitation, so that it must come under the equivalence principle \cite{Nottale1996B,Agnese1997}, i.e., it is independent of the mass of the test-particle. In the Kepler central potential case ($\phi=-GMm/r$), $GM$ provides the natural length-unit of the system under consideration. As a consequence, the parameter ${\cal D}$ reads:
\begin{equation}
\label{C}
{\cal D}=\frac{GM}{2 w},
\end{equation}
where $w$ is a constant that has the dimension of a velocity. The ratio $\alpha_{g}=w/c$ actually plays the role of a macroscopic gravitational coupling constant \cite{Agnese1997,Nottale2000C}.

\subsubsection{Formation and evolution of structures}

Let us now compare our approach with the standard theory of gravitational structure formation and evolution.  By separating the real and imaginary parts of the Schr\"odinger equation we obtain, after a new change of variables, respectively a generalized Euler-Newton equation and a continuity equation, namely,

\begin{equation}
\label{AA1}
m \, (\frac{\partial}{\partial t} + V \cdot \nabla) V  = -\nabla (\phi+Q),\;\;\;\;\;\frac{\partial P}{\partial t} + {\rm div}(P V) = 0,
\end{equation}
where $V$ is the real part of the complex velocity field $\cal V$. In the case when the density of probability is proportional to the density of matter, $P \propto \rho$, this system of equations is equivalent to the classical one used in the standard approach of gravitational structure formation, except for the appearance of an extra potential energy term $Q$ that writes:
\begin{equation}
\label{Q}
Q =-2m{\cal D}^2 \frac{\Delta \sqrt{P}}{\sqrt{P}}. 
\end{equation}

The existence of this potential energy, (which amount to the Bohm potential in standard quantum mechanics) is, in our approach, readily demonstrated and understood: namely, it is the very manifestation of the fractality of space, in similarity with Newton's potential being a manifestation of curvature. We have suggested \cite{Nottale2001A,Nottale2004BB,Nottale2006B} that it could be the origin of the various effects which are usually attributed to an unseen, `dark' matter.

In the case when actual particles achieve the probability density distribution (structure formation), we have $\rho=m_{0} P$; then the Poisson equation (i.e., the field equation) becomes $ \Delta \phi = 4 \pi G m m_{0} |\psi |^{2}$ and it is therefore strongly interconnected with the Schr\"odinger equation (which is here a new form for the equation of motion). 
Such a system of equations is similar to that encountered in the description of superconductivity (Hartree equation). We expect its solutions to provide us with general theoretical predictions for the structures (in position and velocity space) of self-gravitating systems at multiple scales \cite{Nottale1997A,daRocha2003}. This expectation is already supported by the observed agreement of several of these solutions with astrophysical observational data \cite{Nottale1993,Nottale1996B,Nottale2000C,Nottale1997C,Nottale1998E,Nottale1998A,Nottale1998B,Hermann1998}.

\subsubsection{Planetary systems}
\label{sec584}

Let us briefly consider the application of the theory to the formation of planetary systems.
The standard model of formation of planetary systems can be reconsidered in terms of a fractal description of the motion of planetesimals in the protoplanetary nebula. On length-scales much larger than their mean free path, we have assumed  \cite{Nottale1993} that their highly chaotic motion satisfy the three conditions upon which the derivation of a Schr\"odinger equation is based (large number of trajectories, fractality and time symmetry breaking). In modern terms, our proposal is but a `migration' theory, since it amounts to take into account the coupling between planetesimals (or proto-planets) and the remaining disk. But, instead of considering a mean field coupling, we consider the effect of the closest bodies to be the main one, leading to Brownian motion and irreversibility.
 
This description applies to the distribution of planetesimals in the proto-planetary nebula at several embedded levels of hierarchy. Each hierarchical level ($k$) is characterized by a length-scale defining the parameter ${\cal D}_k$  (and therefore the velocity $w_k$) that appears in the generalized Schr\"odinger equation describing this sub-system.
This hierarchical model has allowed us to recover the mass distribution of planets and small planets in the inner and outer solar systems \cite{Nottale1997C}. It is generally supported by the structure of our own solar system, which is made of several subsystems embedded one in another, namely:

\paragraph{The Sun} Through Kepler's third law, the velocity $w=3 \times 144.7=434.1$ km/s is very closely the Keplerian velocity at the Sun radius ($R_\odot=0.00465$ AU corresponds to $w_\odot=437.1$ km/s). 
Moreover, one can also apply our approach to the organization of the sun surface itself. One expect the distribution of the various relevant physical quantities that characterize the solar activity at the Sun surface (sun spot number, magnetic field, etc...) to be described by a wave function whose stationary solutions read $\psi=\psi_0 \; e^{i E t /2 m {\cal D}}$.

The energy $E$ results from the rotational velocity and, to be complete, should also include the turbulent velocity, so that $E=(v_\text{rot}^2+v_\text{turb}^2)/2$. This means that we expect the solar surface activity to be subjected to a fundamental period:
\begin{equation}
\tau=\frac{2 \pi m {\cal D}}{E}=\frac{4 \pi  {\cal D}}{v_\text{rot}^2+v_\text{turb}^2},
\end{equation}
The parameter $\cal D$ at the Sun radius is  ${\cal D}=GM_{\odot}/2w_{\odot}$, then we obtain:
\begin{equation}
 \tau=\frac{2 \pi G  M_{\odot}}{w_{\odot}(v_\text{rot}^2+v_\text{turb}^2)}.
\end{equation}
The average sideral rotation period of the Sun is 25.38 days, yielding a velocity of 2.01 km/s at equator \cite{Pecker1959}. The turbulent velocity has been found to be $v_\text{turb}=1.4\pm0.2$ km/s \cite{Lang1980}. Therefore we find numerically
\begin{equation}
 \tau=(10.2 \pm 1.0) \;{\rm yrs}.
\end{equation}
The observed value of the period of the Solar activity cycle,  $\tau_\text{obs}=11.0$ yrs, nicely supports this theoretical prediction. This is an interesting result, owing to the fact that there is, up to now, no existing theoretical prediction of the value of the solar cycle period, except in terms of very rough order of magnitude \cite{Zeldovich1983}.

\begin{figure}[!ht]
\begin{center}
\includegraphics[width=12cm]{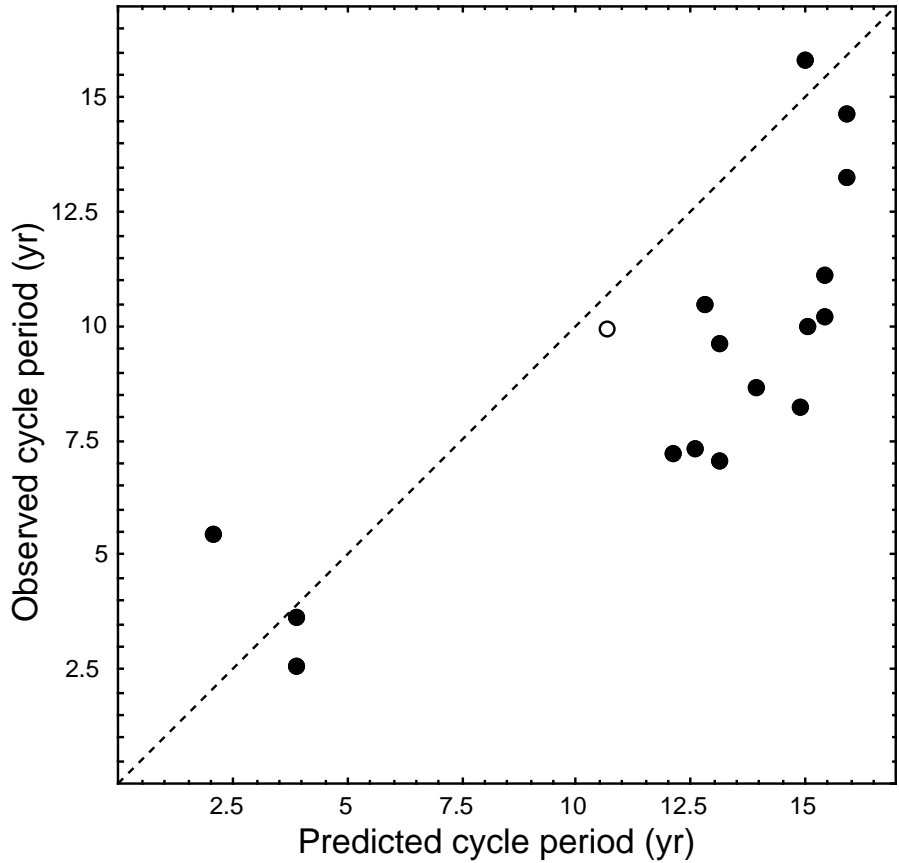}
\caption{\small Comparison between the observed values of the period of solar-like star cycles (inactive stars with better determined behavior in Table 1 of Ref.~\cite{Saar1999}) and the predicted periods (see text). The open point is for the Sun. The correlation is significant at a probability level $P\approx 10^{-4}$ (Student variable $t\approx 5$).}
\label{StarCycle}
\end{center}
\end{figure}

Moreover, since we have now at our disposal a simple and precise formula for a stellar cycle which precisely accounts for the solar period, the advantage is that it can be tested with other stars. The observation of the magnetic activity cycle of distant solar-like stars remains a difficult task, but it has now been performed on several stars. A first attempt gives very encouraging results (see Fig.~\ref{StarCycle}), since we obtain indeed a satisfactory agreement between the observed and predicted periods, in a statistically significant way, despite the small number of objects.

\paragraph{The intramercurial system} organized on the constant $w_\odot=3 \times 144=432$ km/s.
The existence of an intramercurial subsystem is supported by various stable and transient structures observed in dust, asteroid and comet distributions (see \cite{daRocha2003}). We have in particular suggested the existence of a new ring of asteroids, the `Vulcanoid belt', at a preferential distance of about 0.17 AU from the Sun.

\begin{figure}[!ht]
\begin{center}
\includegraphics[width=13cm]{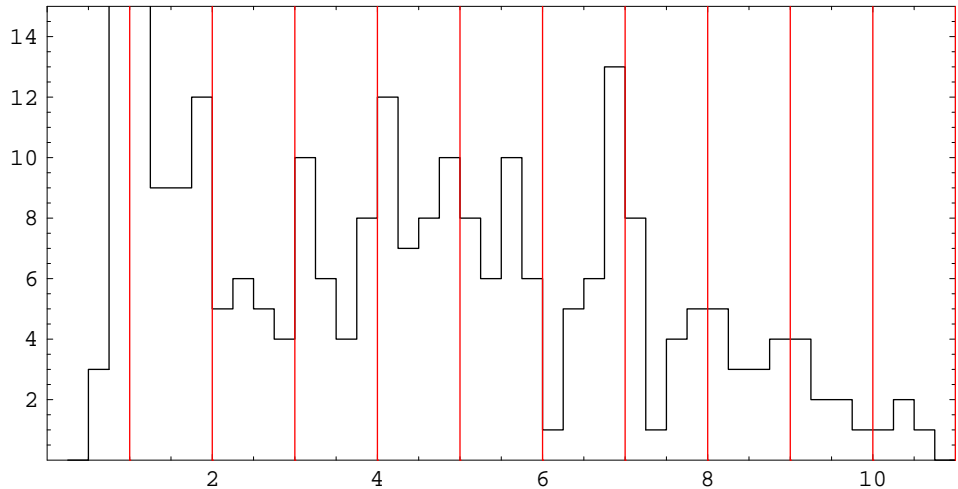}
\caption{\small{Observed distribution of the semi-major axes of 300 exoplanets (June 2008 data \cite{Schneider2008}) and inner solar system planets, compared with the theoretical prediction. The figure gives the histogram of the distribution of the variable $4.83(a/M)^{1/2}$. One predicts the occurence of peaks of probability density for semimajor axes $a_n= GM(n/w_0)^2$, where $n$ is integer, $M$ is the star mass and $w_0=144.7 \pm 0.7 $ km/s is a gravitational coupling constant (see text). The main peak at $n=1$ (fundamental level), which now contains 74 exoplanets, has been cut to better view the secondary peaks. The probability to obtain such an agreement by chance between the predicted (red vertical lines) and observed peaks is now $P= 5 \times 10^{-7}$.}}
\label{fig:exoplanet-a}
\end{center}
\end{figure}

\paragraph{The inner solar system} (earth-like planets), organized with a constant $w_i=144$ km/s (see Fig.~\ref{fig:exoplanet-a}). 

\begin{figure}[!ht]
\begin{center}
\includegraphics[width=16cm]{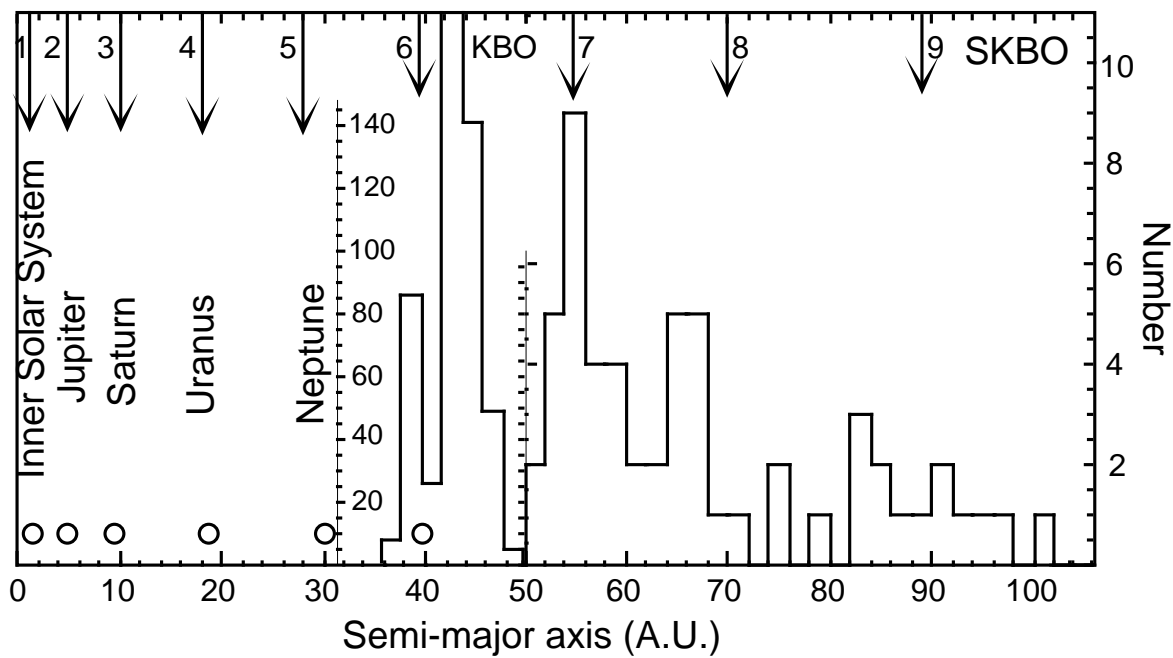}
\caption{\small{Distribution of the semi-major axis of Kuiper belt objects (KBO) and scattered Kuiper belt objects (SKBO), compared with the theoretical predictions (arrows) of probability density peaks for the outer solar system \cite{daRocha2003} (see text). The existence of probability density peaks for the Kuiper belt at $\approx40$, 55, 70, 90 AU, etc...,  has been theoretically predicted in 1993 before the discovery of these objects \cite{Nottale1994C}, and it is now supported by the observational data, in particular by the new small planet Eris at 68 AU, whose mass is larger than Pluto, and which falls close to the expected probability peak $n=8$ at 70 AU (see text).}}
\label{fig:AMSKBO}
\end{center}
\end{figure}

\paragraph{The outer solar system} organized with a constant $w_o=144/5=29$ km/s (see Fig.~\ref{fig:AMSKBO}), as deduced from the fact that the mass peak of the inner solar system lies at the Earth distance ($n=5$). The Jovian giant planets lie from $n=2$ to $n=5$. Pluton lies on $n=6$, but is now considered to be a dwarf planet part of the Kuiper belt.

\paragraph{Kuiper belt}

The recently discovered Kuiper and scattered Kuiper belt objects (Fig. ~\ref{fig:AMSKBO}) show peaks of probability at $n=6$ to 9 \cite{daRocha2003}, as predicted before their discovery \cite{Nottale1994C}. In particular, the predicted peak around 57 AU ($n=7$) is the main observed peak in the SKBO distribution of semi-major axes. The following peak ($n=8$), predicted to be around 70 AU, has received a spectacular verification with the discovery of the dwarf planet Eris (2003 UB313) at 68 AU, whose mass larger than Pluton has recently led to a revision of planetary nomenclature.

\begin{figure}[!ht]
\begin{center}
\includegraphics[width=12cm]{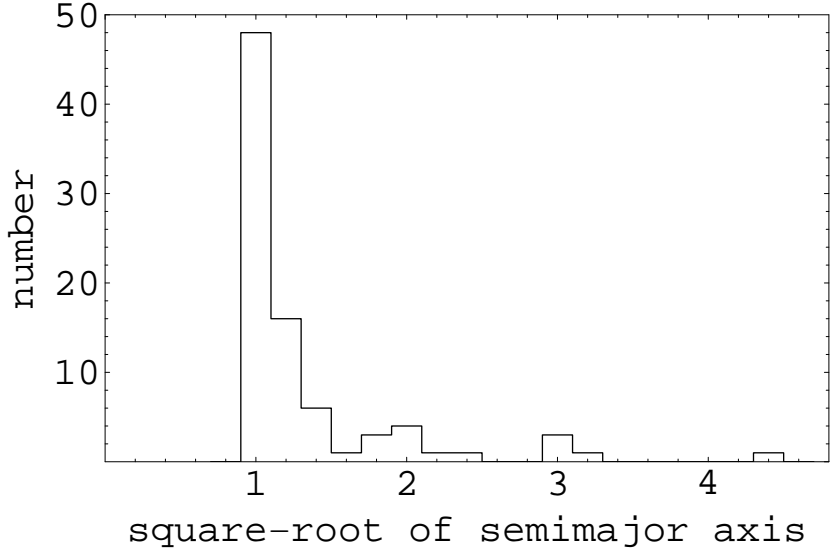}
\caption{\small{Distribution of the semi-major axis of very distant scattered Kuiper belt objects (SKBO) , compared with the theoretical predictions of probability density peaks (see text). We have taken the main SKBO peak at $\approx$57 AU (which is the predicted $n=7$ peak of the outer solar system) as fundamental level ($n=1$) for this new level of hierarchy of the solar system. The figure plots the histogram of the variable $(a/57)^{1/2}$, where $a$ is the semimajor axis of the object orbit in AU. The theoretical prediction, done before the discovery of the distant objects, is that the distribution of this variable should show peaks for integer values, as now verified by the observational data.}}
\label{sedna}
\end{center}
\end{figure}

\paragraph{Distant Kuiper belt}
Beyond these distances, we have been able to predict a new level of hierarchy in the Solar System whose main SKBO peak at 57 AU would be the fundamental level ($n=1$) \cite{Galopeau2004}. The following probability peaks are expected, according to the $n^2$ law, to lie for semi-major axes of 228, 513, 912, 1425, 2052 AU, etc.... Once again this prediction has been validated by the observational data in a remarkable way (see Fig.~\ref{sedna}), since 4 bodies, including the very distant small planet Sedna, have now been discovered in the 513 AU peak ($n=3$), 7 bodies in the 228 AU peak ($n=2$) , and now one very distant object at about 1000 AU (data Minor Planet Center, http://www.cfa.harvard.edu/iau/mpc.html).

\paragraph{Extrasolar planets}

We have suggested more than 16 years ago \cite{Nottale1993,Nottale1994C}, before the discovery of exoplanets, that the theoretical predictions from this approach of planetary formation should apply to all planetary systems, not only our own solar system. Meanwhile more than 300 exoplanets have now been discovered, and the observational data support this prediction in a highly statistically significant way (see \cite{Nottale1996B,Nottale2000C,daRocha2003} and Fig.~\ref{fig:exoplanet-a}).

The presently known exoplanets mainly correspond to the intramercurial and inner solar systems. The theoretical prediction, made in 1993 \cite[Chap. 7.2]{Nottale1993}, according to which the distribution of semi-major axes $a$ is expected to show peaks of probability for integer values of the variable $4.83 (a/M)^{1/2}$, where $M$ is the star mass, remains validated with a high statistical significance (see Fig.~\ref{fig:exoplanet-a}). In particular, in addition to the peaks of probability corresponding to the inner solar system planets ($n=3$ Mercury, $n=4$ Venus, $n=5$ Earth, $n=6$ Mars), two additional predicted peaks of probability, the `fundamental' one at 0.043 AU/M$_\odot$ and the second one at 0.17 AU/M$_\odot$, have been made manifest in extrasolar planetary systems. In particular, the validation of the principal prediction of the SR approach, namely, the main peak at the fundamental level $n=1$, is striking since it now contains more than 80 exoplanets.

\subsection{Applications to sciences of life}\label{cap1}

The scale relativity theory has also been recently applied to sciences other than physical sciences, including sciences of life, sciences of societies, historical \cite{Forriez2005} and geographical sciences \cite{Martin2006,Forriez2006,Forriez2007} and human sciences \cite{Timar2002,Nottale2002D,Nottale2006D,Nottale2008A}. We refer the interested reader to the books \cite{Nottale2000B,Nottale2007I}, to parts of review papers or books  \cite{Nottale2001B,Nottale2004B,Nottale2007G} and full review papers on this specific subject \cite{Auffray2007A,Nottale2007B} for more details.

\subsubsection{Applications of log-periodic laws}

\paragraph{Species evolution}

Let us first consider the application of log-periodic laws to the description of critical time evolution. Recall that a log-periodic generalization to scale invariance has been obtained as a solution to wave-like differential scale equations. Interpreted as a distribution of probability, such solutions therefore lead to a geometric law of progression of probability peaks for the occurence of events.

\begin{figure}[!ht]
\begin{center}
\includegraphics[width=9cm]{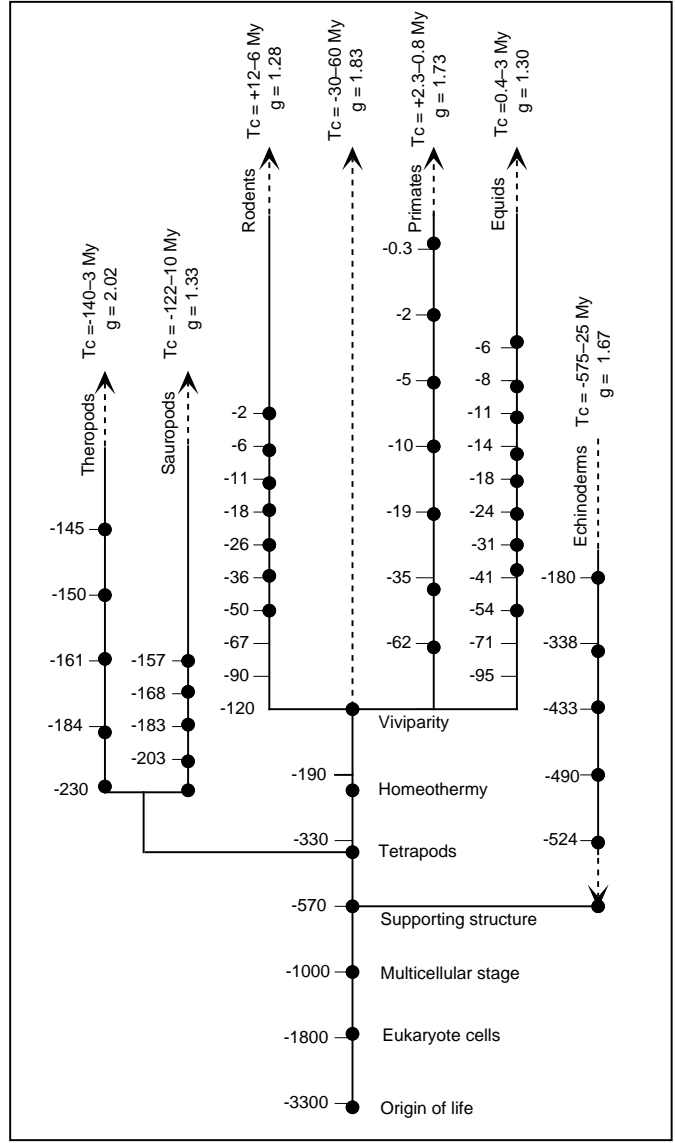}
\caption[Log-periodic evolution on the `tree of life']{\small{The dates of major evolutionary events of seven lineages (common evolution from life origin to viviparity, Theropod and Sauropod dinosaurs, Rodents, Equidae, Primates including Hominidae, and Echinoderms) are plotted as black points in terms of $\log(T _{c}-T)$, and compared with the numerical values from their corresponding log-periodic models (computed with their best-fit parameters). The adjusted critical time $T _{c}$  and scale ratio $g$ are indicated for each lineage \cite{Chaline1999,Nottale2000B,Nottale2002B}.}}
\label{biomedfig1}
\end{center}
\end{figure}

Now several studies have shown that many biological, natural, sociological and economic phenomena obey a log-periodic law of time evolution such as can be found in some critical phenomena : earthquakes \cite{Sornette1995}, stock market crashes  \cite{Sornette1996}, evolutionary leaps  \cite{Chaline1999,Nottale2000B,Nottale2002B}, long time scale evolution of western and other civilizations \cite{Nottale2000B,Nottale2002B,Grou2004}, world economy indices dynamics \cite{Johansen2001}, embryogenesis \cite{Cash2002}, etc... Thus emerges the idea that this behaviour typical of temporal crisis could be extremely widespread, as much in the organic world as in the inorganic one \cite{Sornette1998}. 

In the case of species evolution, one observes the occurrence of major evolutionary leaps leading to bifurcations among species, which proves the existence of punctuated evolution \cite{Gould1977} in addition to the gradual one. The global pattern is assimilated to a `tree of life', whose bifurcations are identified to evolutionary leaps, and branch lengths to the time intervals between these major events \cite{Chaline1999}. As early recognized by Leonardo da Vinci, the branching of vegetal trees and rivers may be described as a first self-similar approximation by simply writing that the ratio of the lengths of two adjacent levels is constant in the mean. We have made a similar hypothesis for the time intervals between evolutionary leaps, namely, $(T_n-T_{n-1})/(T_{n+1}-T_{n})=g$. Such a geometric progression yields a log-periodic acceleration for $g>1$, a deceleration for $g<1$, and a periodicity for $g=1$. Except when $g=1$, the events converge toward a critical time $T_c$ which can then be taken as reference, yielding the following law for the event $T_n$ in terms of the rank $n$:
\beq
T _{n} = T _{c} + (T _{0} - T _{c})\, g ^{-n},
\eeq
where $T _{0}$ is any event in the lineage, $n$ the rank of occurrence of a given event and $g$ is the scale ratio between successive time intervals. Such a chronology is periodic in terms of logarithmic variables, i.e., $\log|T _{n} - T _{c}| = \log|T _{0} - T _{c}| - n \log g$. 

This law is dependent on two parameters only, $g$ and $T _{c}$, which of course have no reason a priori to be constant for the entire tree of life. Note that $g$ is not expected to be an absolute parameter, since it depends on the density of events chosen, i.e., on the adopted threshhold in the choice of their importance (namely, if the number of events is doubled, $g$ is replaced by $\sqrt{g}$). Only a maximal value of $g$, corresponding to the very major events, could possibly have a meaning. On the contrary, the value of $T _{c}$ is expected to be a characteristic of a given lineage, and therefore not to depend on such a choice. This expectation is supported by an analysis of the fossil record data under various choices of the threshold on the events, which have yielded the same values of $T_c$ within error bars \cite{Nottale2002B}.  

A statistically significant log-periodic acceleration has been found at various scales for global life evolution, for primates, for sauropod and theropod dinosaurs, for rodents and North American equids. A deceleration law was conversely found in a statistically significant way for echinoderms and for the first steps of rodents evolution (see Fig.~\ref{biomedfig1} and more detail in Refs.~\cite{Chaline1999,Nottale2000B,Nottale2002B}). One finds either an acceleration toward a critical date $T_c$ or a deceleration from a critical date, depending on the considered lineage. 

It must be remarked that the observed dates follow a log-periodic law only in the mean, and show a dispersion around this mean (see \cite[p.~320]{Nottale2000B}. In other words, this is a statistical acceleration or deceleration, so that the most plausible interpretation is that the discrete $T_n$ values are nothing but the dates of peaks in a continuous probability distribution of the events. Moreover, it must also be emphasized that this result does not put the average constancy of the mutation rate in question. This is demonstrated by a study of the cytochrome c tree of branching (in preparation), which is based on genetic distances instead of geological chronology, and which nevertheless yields the same result, namely, a log-periodic acceleration of most lineages, and a periodicity (which corresponds to a critical time tending to infinity) in some cases. The average mutation rate remains around 1/20 Myr since about 1 Gyr, so that one cannot escape the conclusion that the number of mutations needed to obtain a major evolutionary leap decreases with time among many lineages, and increases for some of them.

\paragraph{Embryogenesis and human development}
Considering the relationships between phylogeny and ontogeny, it appeared interesting to verify whether the log-periodic law describing the chronology of several lineages of species evolution may also be applied to the various stages in human embryological development. The result, (see Figure in Chaline's contribution), is that a statistically significant log-periodic deceleration with a scale ratio  $g=1.71 \pm 0.01$ is indeed observed, starting from a critical date that is consitent with the conception date \cite{Cash2002}.

\paragraph{Evolution of societies}\label{atteo}

Many observers have commented on the way historical events accelerate. Grou \cite{Grou1987} has shown that the economic evolution since the neolithic can be described in terms of various dominating poles which are submitted to an accelerating crisis-nocrisis pattern, which has subsequently been quantitatively analysed using log-periodic laws. 

\begin{figure}[!ht]
\begin{center}
\includegraphics[width=16cm]{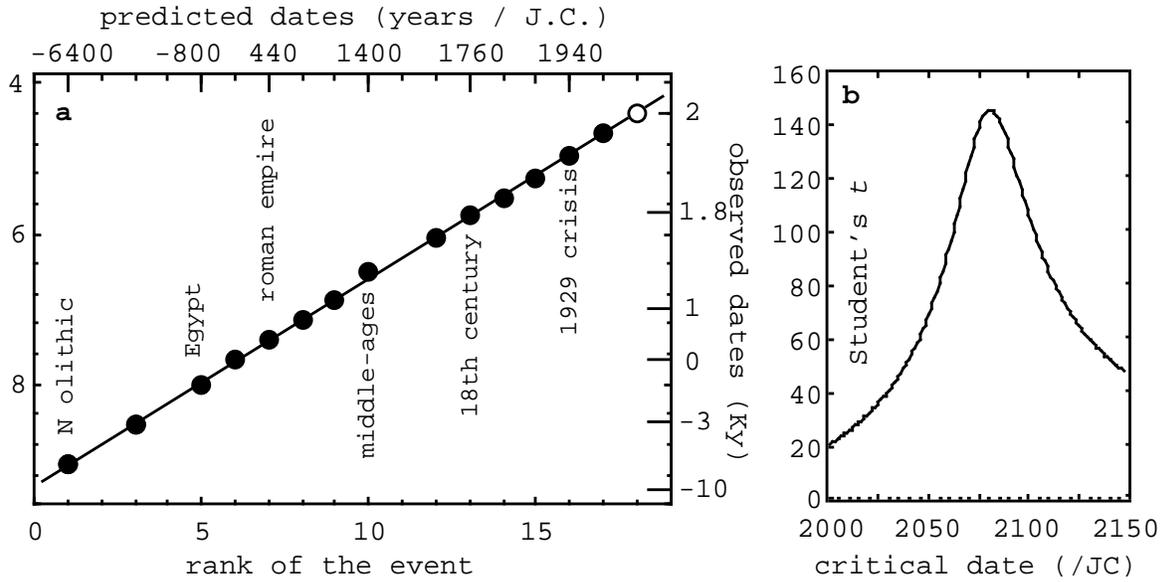}
\caption[Log-periodic evolution of economic crises]{\small{Comparison of the median dates of the main economic crises of western civilization with a log-periodic accelerating law of critical date $T _{c }= 2080$ and scale ratio $g  = 1.32$ (figure a). The last white point corresponds to the predicted next crisis (1997-2000) at the date of the study (1996), as has been later supported in particular by the 1998 and 2000 market crashes, while the next crises are now predicted for (2015-2020), then (2030-2035). Figure b shows the estimation of the critical date through the optimisation of the Student's $t$ variable. This result is statistically significant, since the probability to obtain such a high peak by chance is $P<10 ^{-4}$.}}
\label{biomedfig4}
\end{center}
\end{figure}

For the Western civilization since the Neolithic (i.e., on a time scale of about 8000 years), one finds that a log-periodic acceleration with scale factor $g = 1.32 \pm  0.018$ occurs toward $T _{c} = 2080 \pm  30$ (see Fig.~\ref{biomedfig4}), in a statistically highly significant way.
This result has been later confirmed by Johansen and Sornette \cite{Johansen2001} by an independent study on various market, domestic, research and development, etc... indices on a time scale of about 200 years, completed by demography on a time scale of about 2000 years. They find critical dates for these various indices in the range 2050-2070, which support the longer time scale result. 

One of the intriguing features of all these results is the frequent occurence of values of the scale ratio $g$ close to $g=1.73$ and its square root $1.32$ (recall that one passes from a value of $g$ to its square root by simply doubling the number of events). This suggests once again a discretization of the values of this scale ratio, that may be the result of a probability law (in scale space) showing quantized probability peaks . We have considered the possibility that $g=1.73 \approx \sqrt{3}$ could be linked to a most probable branching ratio of 3 \cite{Chaline1999,Nottale2000B}, while Queiros-Cond\'e \cite{Queiros2000} has proposed a `fractal skin' model for understanding this value.

\subsubsection{History and geography}
The application of the various tools and methods of the scale relativity theory to history and geography has been proposed by Martin and Forriez \cite{Forriez2005,Martin2006,Forriez2006,Forriez2007}.  Forriez has shown that the chronology of some historical events (various steps of evolution of a given site) recovered from archeological and historical studies can be fitted by a log-periodic deceleration law with again $g \approx 1.7$ and a retroprediction of the foundation date of the site from the critical date \cite{Forriez2005,Forriez2006}. Moreover, the various differential equation tools developed in the scale  relativity approach both in scale and position space, including the nonlinear cases of variable fractal dimensions, have been found to be particularly well adapted to the solution of geographical problems \cite{Martin2006}.

 \subsubsection{Predictivity}
 Although these studies remain, at that stage, of an empirical nature (it is only a purely chronological analysis which does not take into account the nature of the events), they nevertheless provide us with a beginning of predictivity. Indeed, the fitting law is a two parameter function ($T_c$ and $g$) that is applied to time intervals, so that only three events are needed to define these parameters. Therefore the subsequent dates are predicted after the third one, in a statistical way. Namely, as already remarked, the predicted dates should be interpreted as the dates of the peaks of probability for an event to happen. 

Examples of such a predictivity (or retropredictivity) are: 
  
 \noindent (i) the retroprediction that the common Homo-Pan-Gorilla ancestor (expected, e.g., from genetic distances and phylogenetic studies), has a more probable date of appearance at $\approx -10$ millions years \cite{Chaline1999}; its fossil has not yet been discovered (this is one of the few remaining `missing links'); 
 
 \noindent (ii) the prediction of a critical date for the long term evolution of human societies around the years 2050-2080 \cite{Nottale2000B,Johansen2001,Nottale2002B,Grou2004}; 
 
 \noindent (iii) the finding that the critical dates of rodents may reach $+60$ Myrs in the future, showing their large capacity of evolution, in agreement with their known high biodiversity; 
 
 \noindent (iv) the finding that the critical dates of dinosaurs are about $-150$ Myrs in the past, indicating that they had reached the end of their capacity of evolution (at least for the specific morphological characters studied) well before their extinction at $-65$ Myrs; 
 
 \noindent (v) the finding that the critical dates of North american Equids is, within uncertainties, consistent with the date of their extinction, which may mean that, contrarily to the dinosaur case, the end of their capacity of evolution has occured during a phase of environmental change that they have not been able to deal with by the mutation-selection process;
 
 \noindent (vi) the finding that the critical date of echinoderms (which decelerate instead of accelerating) is, within uncertainties, the same as that of their apparition during the PreCambrian-Cambrian radiation, this supporting the view of the subsequent events as a kind of ``scale wave" expanding from this first shock.

\begin{figure}[!ht]
\begin{center}
\includegraphics[width=14cm]{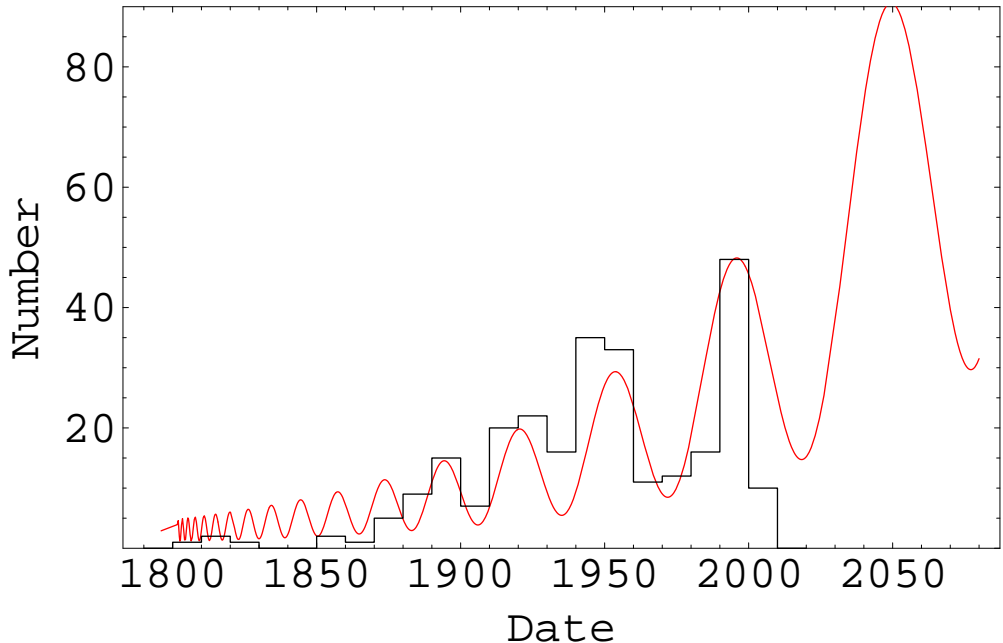}
\caption{\small{Observed rate of Southern California earhquakes of magnitude larger than 5 (histogram). The data are taken from the U.S. Geological Survey EarthQuake Data Center (years 1932-2006) and EarthQuake Data Base  (Historical earthquakes, years 1500-1932). This rate is well fitted by a power law subjected to a log-periodic fluctuation decelerating since a critical date $T_c=1796$ (red fluctuating line). The model predicts the next probability peak around the years 2050 \cite{Nottale2007H}.}}
\label{califearthquake}
\end{center}
\end{figure}

 \subsubsection{Applications in Earth sciences}
As last examples of such a predictivity, let us give some examples of applications of critical laws (power laws in $|T-T_c|^\gamma$ and their log-perodic generalizations) to problems encountered in Earth sciences, namely, earthquakes (California and Sichuan) and decline of Arctic sea ice.

\begin{figure}[!ht]
\begin{center}
\includegraphics[width=10cm]{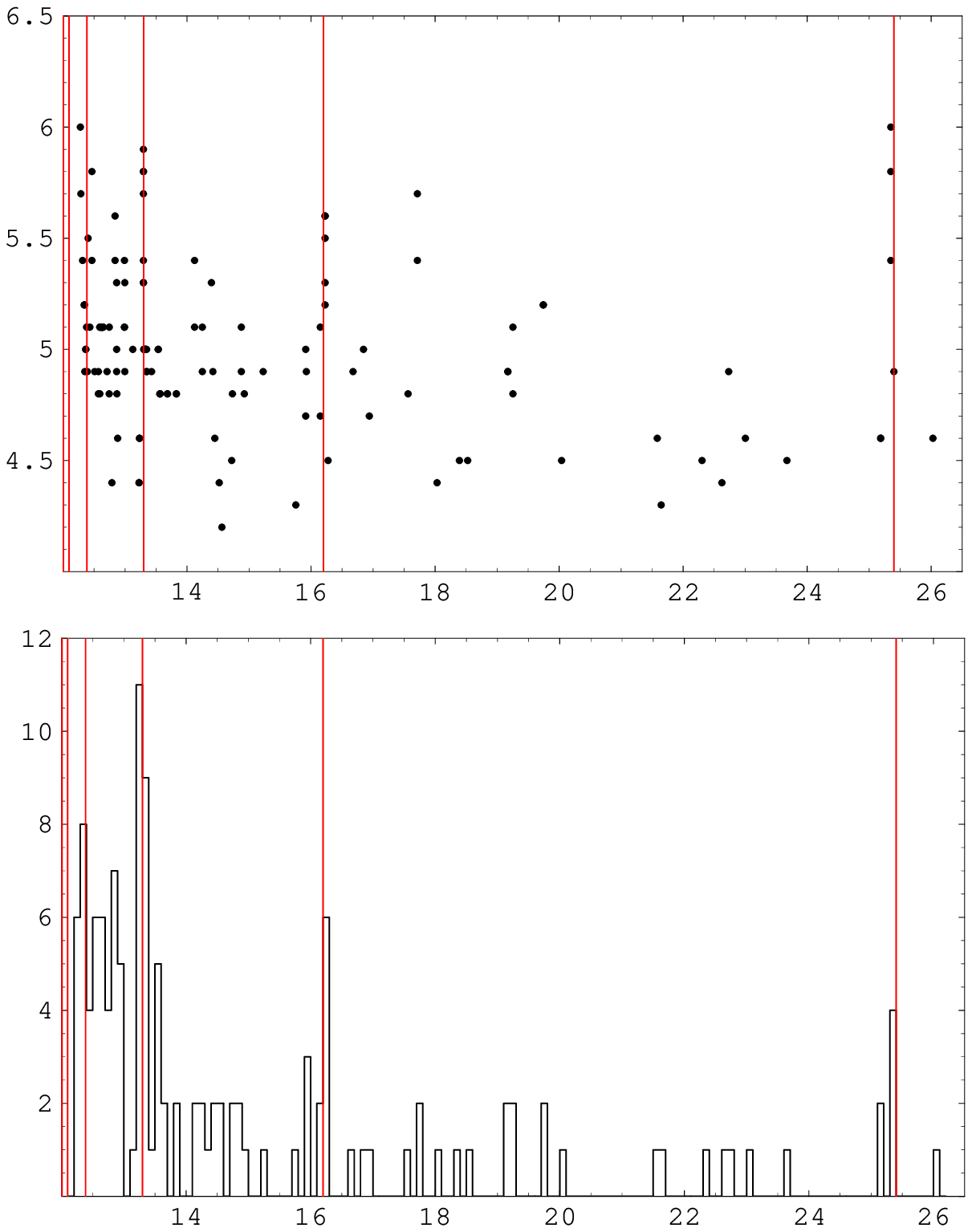}
\caption{\small{Time evolution during 14 days of the replicas of the May 12,  2008 Sichuan earthquake (data obtained and studied May 27, 2008 from the seismic data bank EduSeis Explorer, 
http://aster.unice.fr/EduSeisExplorer/form-sis.asp). The (up) figure gives the magnitudes of the replicas and the (down) figure the rate of replicas. Both show a continuous decrease to which are added discrete sharp peaks. The peaks which are common to both diagrams show a clear deceleration according to a log-periodic law starting from the main earthquake (red vertical lines), which allows one to predict the next strongest replicas with a good precision. For example, the peak of replicas of 25 May 2008 could be predicted  with a precision of 1.5 day from the previous peaks. Reversely, the date of the main earthquake (May 12.27 2008, magnitude 7.9) can be retropredicted from that of the replicas with a precision of 6 h.}}
\label{Sichuan}
\end{center}
\end{figure}

\paragraph{California earthquakes}

The study of earthquakes has been one of the first domain of application of critical and log-periodic laws \cite{Sornette1995,Allegre1982}. The rate of California earthquakes is found to show a very marked log-periodic deceleration \cite{Nottale2007H,Nottale2007I}. We show indeed in Fig.~\ref{califearthquake} the observed rate of Southern California earhquakes of magnitude larger than 5, compared with a log-periodic deceleration law. This model allows us to predict future peaks of probability around the years 2050 then 2115.
 
\begin{figure}[!ht]
\begin{center}
\includegraphics[width=14cm]{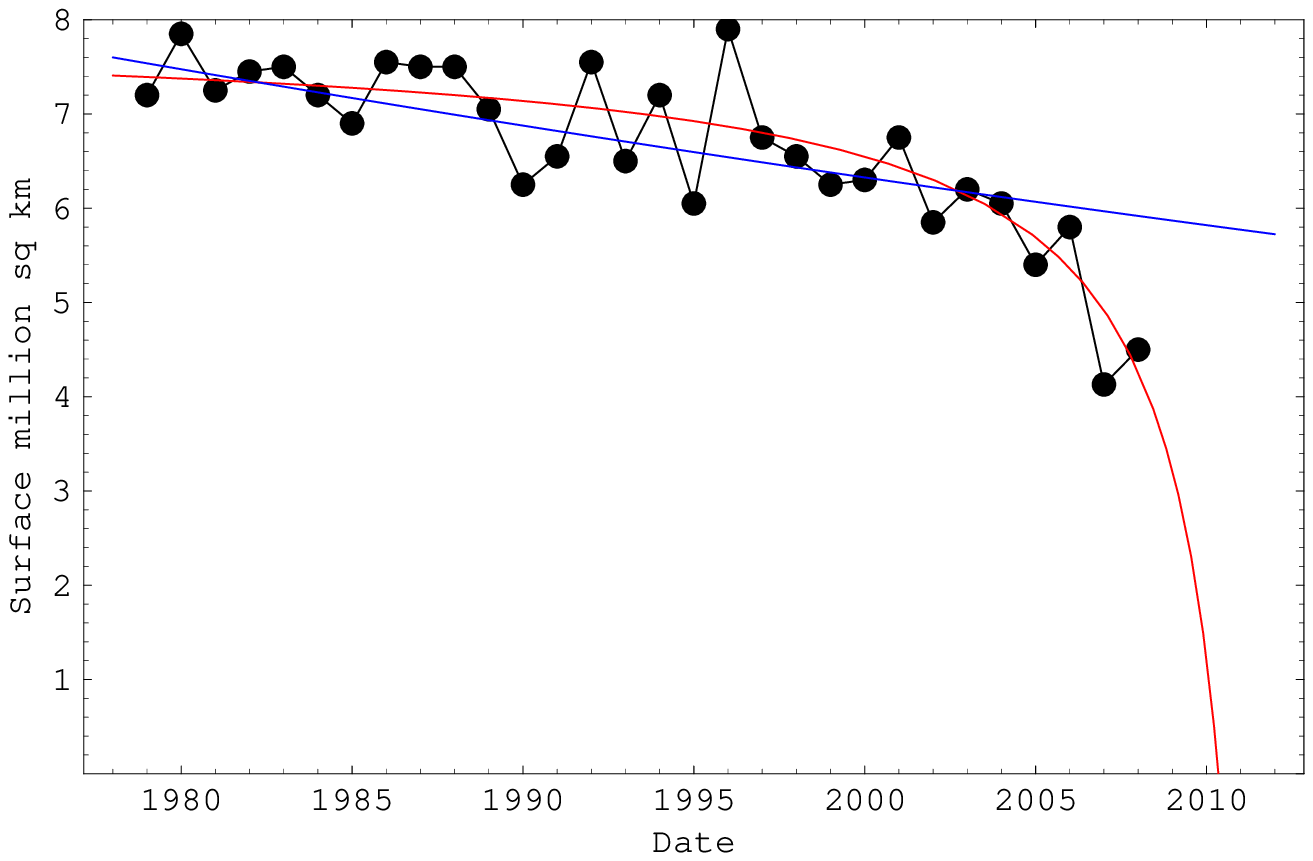}
\caption{\small{Observed evolution of the minimum arctic sea ice extent, according to the data of the U.S. National Snow and Ice Data Center (NSIDC, http://nsidc.org/), from 1979 to 2008. The minimum ocurs around 15 September of each year. This evolution is compared to: (i) the standard fit corresponding to an assumed constant rate of extent decrease (blue line); (ii) a fit by a critical law accelerating toward a critical date $T_c=2012$. The second fit is far better and has allowed us to predict the 2007 and 2008 low points before their observation \cite{Nottale2007L}. It implies that the arctic sea is expected to be totally free from ice by September 2011.}}
\label{Arctic1}
\end{center}
\end{figure}

\paragraph{Sichuan 2008 earthquake}

The May 2008 Sichuan earthquake and its replicas also yields a good example of log-periodic deceleration, but on a much smaller time scale (see Fig~\ref{Sichuan}).

\paragraph{Arctic sea ice extent}

It is now well-known that the decrease of arctic sea-ice extent has shown a strong acceleration in 2007 and 2008 with respect to the current models assuming a constant rate ($\approx 8 \%$ by decade), which predicted in 2006 a total disappearance of the ice at minimum (15 september) for the end of the century.  From the view point of these models, the 2007 and now 2008 values (see Fig.~\ref{Arctic1}) were totally unexpected. 

However, we have proposed, before the knowledge of the 2007 minimum, to fit the data with a critical law of the kind $y=y_0 - a |T-T_c|^\gamma$. Such an accelerating law has the advantage to include in its structure the fact that one expect the ice to fully disappear after some date, while the constant rate law formally pushed the date of total disappearance to infinity. The fit of the data up to 2006 with the critical law was already far better than with the constant rate law, and it actually allowed us to predict a full disappearance epoch far closer than previously expected and a low 2007 point \cite{Nottale2007L}. The 2008 point has confirmed the validity of the model in an impressive way (Fig.~\ref{Arctic1}). We obtain by the $\chi^2$ method a best fit for the minimum  ice surface (in square kilometers), $y=8-12.3 \times |T-2012|^{-0.86}$. The critical time is as early as $T_c=2012$, which means that a full ice melting is predicted for September 2011, and is even possible for September 2010, for which the model gives only 1.2  million km$^2$ of remaining ice surface.

The application of the same method to the mean surface data during August and October months also shows a clear acceleration toward $T_c=2013$, which means that only one year later (2012) the arctic sea can be expected to be free from ice during several months (August to October).

\subsection{Applications of scale relativity to biology} 
One may consider several applications to biology of the various tools and methods of the scale relativity theory, namely, generalized scale laws, macroscopic quantum-type theory and Schr\"odinger equation in position space then in scale space and emergence of gauge-type fields and their associated charges from fractal geometry \cite{Nottale1993,Nottale2000B,Nottale2004A,Auffray2007A,Nottale2007A}. One knows that biology is founded on biochemistry, which is itself based on thermodynamics, to which we contemplate the future possibility to apply the macroquantization tools described in the theoretical part of this article. Another example of future possible applications is to the description of the growth of polymer chains, which could have consequences for our understanding of the nature of DNA and RNA molecules.

Let us give some explicit examples of such applications.

\subsection{Confinement} The solutions of non-linear scale equations such as that involving a harmonic oscillator-like scale force \cite{Nottale1997B} may be meaningful for biological systems. Indeed, its main feature is its capacity to describe a system in which a clear separation has emerged between an inner and an outer region, which is one of the properties of the first prokaryotic cell. We have seen that the effect of a scale harmonic oscillator force results in a confinement of the large scale material in such a way that the small scales may remain unaffected. 

Another interpretation of this scale behavior amounts to identify the zone where the fractal dimension diverges (which corresponds to an increased `thickness` of the material) as the description of a membrane. It is indeed the very nature of biological systems to have not only a well-defined size and a well-defined separation between interior and exterior, but also systematically an interface between them, such as membranes or walls. This is already true of the simplest prokaryote living cells. Therefore this result suggests the possibility that there could exist a connection between the existence of a scale field (e.g., a global pulsation of the system, etc..) both with the confinement of the cellular material and with the appearance of a limiting membrane or wall \cite{Nottale2004A}. This is reminiscent of eukaryotic cellular division which involves both a dissolution of the nucleus membrane and a deconfinement of the nucleus material, transforming, before the division, an eukaryote into a prokaryote-like cell. This could be a key toward a better understanding of the first major evolutionary leap after the appearance of cells, namely the emergence of eukaryotes.


\subsection{Morphogenesis}
The generalized Schr\"odinger equation (in which the Planck constant $\hbar$ can be replaced by a macroscopic constant) can be viewed as a fundamental equation of morphogenesis. It has not been yet considered as such, because its unique domain of application was, up to now, the microscopic (molecular, atomic, nuclear and elementary particle) domain, in which the available information was mainly about energy and momentum.

However, scale relativity extends the potential domain of application of Schr\"odinger-like equations to every systems in which the three conditions (infinite or very large number of trajectories, fractal dimension of individual trajectories, local irreversibility) are fulfilled. Macroscopic Schr\"odinger equations can be constructed, which are not based on Planck's constant $\hbar$, but on constants that are specific of each system (and may emerge from their self-organization). 

Now the three above conditions seems to be particularly well adapted to the description of living systems. Let us give a simple example of such an application. 

In living systems, morphologies are acquired through growth processes. One can attempt to describe such a growth in terms of an infinite family of virtual, fractal and locally irreversible, trajectories. Their equation can therefore be written under the form of a fractal geodesic equation, then it can be integrated as a Schr\"odinger equation.

If one now looks for solutions describing a growth from a center, one finds that this problem is formally identical to the problem of the formation of planetary nebulae \cite{daRocha2003}, and, from the quantum point of view, to the problem of particle scattering, e.g., on an atom. The solutions looked for correspond to the case of the outgoing spherical probability wave. 

Depending on the potential, on the boundary conditions and on the symmetry conditions, a large family of solutions can be obtained. Considering here only the simplest ones, i.e., central potential and spherical symmetry, the probability density distribution of the various possible values of the angles are given in this case by the spherical harmonics,
\beq
P(\theta,\varphi) =|Y_{lm}(\theta,\varphi)|^2.
\eeq
These functions show peaks of probability for some angles, depending on the quantized values of the square of angular momentum $L^2$ (measured by the quantum number $l$) and of its projection $L_z$ on axis $z$ (measured by the quantum number $m$).

\begin{figure}[!ht]
\begin{center}
\includegraphics[width=7cm]{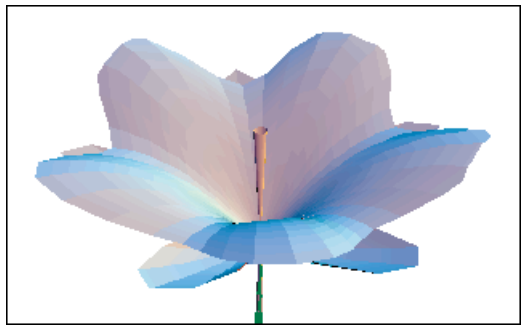}
\caption{\small{Morphogenesis of a `flower'-like structure, solution of a generalized Schr\"odinger equation that describes a growth process from a center($l=5,\; m=0$). The `petals', `sepals' and `stamen' are traced along angles of maximal probability density. A constant force of `tension' has been added, involving an additional curvature of `petals', and a quantization of the angle $\theta$ that gives an integer number of `petals' (here, $k=5$).}}
\label{fig:fleur}
\end{center}
\end{figure}

Finally a more probable morphology is obtained by `sending' matter along angles of maximal probability. The biological constraints leads one to skip to cylindrical symmetry. This yields in the simplest case a periodic quantization of the angle $\theta$ (measured by an additional quantum number $k$), that gives rise to a separation of discretized `petals'. Moreover there is a discrete symmetry breaking along the $z$ axis linked to orientation (separation of `up' and `down' due to gravity, growth from a stem). The solutions obtained in this way show floral `tulip'-like shapes (see Fig.~\ref{fig:fleur} and \cite{Nottale2001B,Nottale2004A,Nottale2007A}). 

Coming back to the foundation of the theory, it is remarkable that these shapes are solutions of a geodesic, strongly covariant equation $\dfr {\cal V} /dt=0$, which has the form of the Galilean motion equation in vacuum in the absence of external force. Even more profoundly, this equation does not describe the motion of a particle, but purely geometric virtual possible paths, this given rise to a description in terms of a probability density which plays the role of a potential for the real particle (if any, since, in the application to elementary particles, we identify the `particles' with the geodesics themselves, i.e., they become pure relative geometric entities devoid of any proper existence).

\subsection{Origin of life}
The problems of origin are in general more complex than the problems of evolution. Strictly, there is no `origin' and both problems could appear to be similar, since the scientific and causal view amounts to consider that any given system finds its origin in an evolution process. However, systems are in general said to evolve if they keep their nature, while the question is posed in terms of origin when a given system appears from another system of a completely different nature, and moreover, often on times scales which are very short with respect to the evolution time. An example in astrophysics is the origin of stars and planetary systems from the interstellar medium, and in biology the probable origin of life from a prebiotic medium. 

A fondamentally new feature of the scale relativity approach concerning such problems  is that the Schr\"odinger form taken by the geodesic equation can be interpreted as a general tendency for systems to which it applies to make structures, i.e., to lead to self-organization. In the framework of a classical deterministic approach, the question of the formation of a system is always posed in terms of initial conditions. In the new framework, the general existence of stationary solutions allows structures to be formed whatever the initial conditions, in correspondence with the field, the symmetries and the boundary conditions (namely the environmental conditions in biology), and in function of the values of the various conservative quantities that characterize the system.

Such an approach could allow one to ask the question of the origin of life in a renewed way. This problem is the analog of the `vacuum' (lowest energy) solutions, i.e., of the passage from a non-structured medium to the simplest, fundamental level structures. In astrophysics and cosmology, the problem amounts to understand the apparition, from the action of gravitation alone, of structures (planets, stars, galaxies, clusters of galaxies, large scale structures of the Universe) from a highly homogeneous and non-structured medium whose relative fluctuations were smaller than $10^{-5}$ at the time of atom formation. In the standard approach to this problem a large quantity of postulated and unobserved dark matter is needed to form structures, and even with this help the result is dissatisfying. In the scale relativity framework, we have suggested that the fundamentally chaotic behavior of particle trajectories leads to an underlying fractal geometry of space, which involves a Schr\"odinger form for the equation of motion, leading both to a natural tendency to form structures and to the emergence of an additional potential energy, identified with the `missing mass(-energy)'. 

The problem of the origin of life, although clearly far more difficult and complex, shows common features with this question. In both cases one needs to understand the apparition of new structures, functions, properties, etc... from a medium which does not yet show such structures and functions. In other words, one need a theory of emergence. We hope that scale relativity is a good candidate for such a theory, since it owns the two required properties: (i) for problems of origin, it gives the conditions under which a weakly structuring or destructuring (e.g., diffusive) classical system may become quantum-like and therefore structured; (ii) for problems of evolution, it makes use of the self-organizing property of the quantum-like theory.

We therefore tentatively suggest a new way to tackle the question of the origin of life (and in parallel, of the present functionning of the intracellular medium) \cite{Nottale2004A,Auffray2007A,Nottale2007B}. The prebiotic medium on the primordial Earth is expected to have become chaotic in such a way that, on time scales long with respect to the chaos time (horizon of predictibility), the conditions that underlie the transformation of the motion equation into a Schr\"odinger-type equation, namely, complete information loss on angles, position and time leading to a fractal dimension 2 behavior on a range of scales reaching a ratio of at least $10^4$-$10^5$, be fulfilled. Since the chemical structures of the prebiotic medium have their lowest scales at the atomic size, this means that, under such a scenario, one expects the first organized units to have appeared at a scale of about $10\; \mu$m, which is indeed a typical scale for the first observed prokaryotic cells. The spontaneous transformation of a classical, possibly diffusive mechanics, into a quantum-like mechanics, with the diffusion coefficient becoming the quantum self-organization parameter $\cal D$ would have immediate dramatic consequences: quantization of energy and energy exchanges and therefore of information, apparition of shapes and quantization of these shapes (the cells can be considered as the `quanta' of life), spontaneous duplication and branching properties (see herebelow), etc... Moreover, due to the existence of a vacuum energy in quantum mechanics (i.e., of a non vanishing minimum energy for a given system), we expect the primordial structures to appear at a given non-zero energy, without any intermediate step.

Such a possibility is supported by the symplectic formal structure of thermodynamics \cite{Peterson1979}, in which the state equations are analogous to Hamilton-Jacobi equations. One can therefore contemplate the possibility of a future `quantization' of thermodynamics, and then of the chemistry of solutions, leading to a new form of macroscopic quantum (bio)-chemistry, which would hold both for the  prebiotic medium at the origin of life and for today's intracellular medium.

In such a framework, the fundamental equation would be the equation of molecular fractal geodesics, which could be transformed into a Schr\"odinger equation for wave functions $\psi$. This equation describes an universal tendency to make structures in terms of a probability density $P$ for chemical products (constructed from the distribution of geodesics), given by the squared modulus of the wave function $\psi= \sqrt{P} \times e^{i \theta}$. Each of the molecules being subjected to this probability (which therefore plays the role of a potentiality), it is proportional to the concentration $c$ for a large number of molecules, $P\propto c$ but it also constrains the motion of individual molecules when they are in small number (this is similar to a particle-by-particle Young slit experiment).

Finally, the Schr\"odinger equation may in its turn be transformed into a continuity and Euler hydrodynamic-like system  (for the velocity $V=(v_+ + v_-)/2$ and the probability $P$) with a quantum potential depending on the concentration when $P\propto c$,
\beq
Q=-2{\cal D}^2 \frac{\Delta \sqrt{c}}{\sqrt{c}}.
\label{ccc}
\eeq
This hydrodynamics-like system also implicitly contains as a sub-part a standard diffusion Fokker-Planck equation with diffusion coefficient $\cal D$ for the velocity $v_+$. It is therefore possible to generalize the standard classical approach of biochemistry which often makes use of fluid equations, with or without diffusion terms (see, e.g., \cite{Noble2002,Smith2004}).  

Under the point of view of this third representation, the spontaneous transformation of a classical system into a quantum-like system through the action of fractality and small time scale irreversibility manifests itself by the appearance of a quantum-type potential energy in addition to the standard classical energy balance. We therefore predict that biological systems must show an additional energy (quite similar to the missing energy of cosmology usually attributed to a never found `dark matter') given by the above relation (\ref{ccc}) in terms of concentrations, when their total measured energy balance is compared to the classically expected one. 

But we have also shown that the opposite of a quantum potential is a diffusion potential. Therefore, in case of simple reversal of the sign of this potential energy, the self-organization properties of this quantum-like behavior would be immediately turned, not only into a weakly organized classical system, but even into an increasing entropy diffusing and desorganized system. We tentatively suggest \cite{Nottale2007A} that such a view may provide a renewed way of approach to the understanding of tumors, which are characterized, among many other features, by both energy affinity and morphological desorganization.

\subsection{Duplication}
\begin{figure}[!ht]
\begin{center}
\includegraphics[width=8cm]{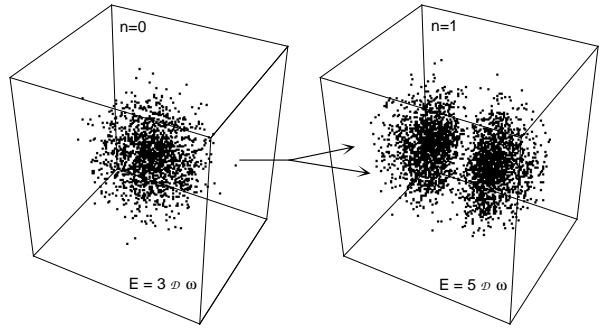}
\caption{\small{Model of duplication. The stationary solutions of the Schr\"odinger equation in a 3D harmonic oscillator potential can take only discretized morphologies in correspondence with the quantized value of the energy. Provided the energy increases from the one-structure case ($E_0=3{\cal D} \omega$), no stable solution can exist before it reaches the second quantized level at $E_1=5{\cal D} \omega$. The solutions of the time-dependent equation show that the system jumps from the one structure to the two-structure morphology.}}
\label{duplication}
\end{center}
\end{figure}

Secondly, the passage from the fundamental level to the first excited level now provides one with a (rough) model of duplication (see Figs.~\ref{duplication} and \ref{bifurcation}). Once again, the quantization implies that, in case of energy increase, the system will not increase its size, but will instead be lead to jump from a single structure to a binary structure, with no stable intermediate step between the two stationary solutions $n=0$ and $n=1$. Moreover, if one comes back to the level of description of individual trajectories, one finds that from each point of the initial one body-structure there exist trajectories that go to the two final structures. In this framework, duplication is expected to be linked to a discretized and precisely fixed jump in energy.

It is clear that, at this stage, such a model is extremely far from describing the complexity of a true cellular division, which it did not intend to do. Its interest is to be a generic and general model for a spontaneous duplication process of quantized structures, linked to energy jumps.  Indeed, the jump from one to two probability peaks when going from the fundamental level to the first excited level is found in many different situations of which the harmonic oscillator case is only an example. Moreover, this duplication property is expected to be conserved under more elaborated versions of the description provided the asymptotic small scale behavior remains of constant fractal dimension $D_F\approx 2$, such as, e.g., in cell wall-like models based on a locally increasing effective fractal dimension.  

\subsection{Bifurcation, branching process}

\begin{figure}[!ht]
\begin{center}
\includegraphics[width=5cm]{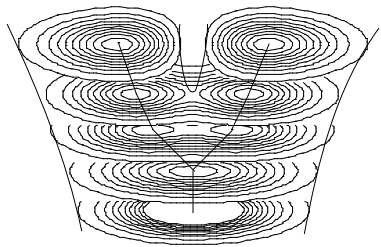}
\caption{\small{Model of branching and bifurcation. Successive solutions of the time-dependent 2D Schr\"odinger equation in an harmonic oscillator potential are plotted as isodensities. The energy varies from the fundamental level ($n=0$) to the first excited level ($n=1$), and as a consequence the system jumps from a one-structure to a two-structure morphology. }}
\label{bifurcation}
\end{center}
\end{figure}

Such a model can also be applied to a first rough description of a branching process (Fig.~\ref{bifurcation}), e.g., in the case of a tree growth when the previous structure remains instead of disappearing as in cell duplication.

Note finally that, although such a model is still clearly too rough to claim that it describes biological systems, it may already be improved by combining with it various other functional and morphological elements which have been obtained. Namely, one may apply the duplication or branching process to a system whose underlying scale laws (which condition the derivation of the generalized Schr\"odinger equation) include (i) the model of membrane through a fractal dimension that becomes variable with the distance to a center; (ii) the model of multiple hierarchical levels of organization depending on `complexergy' (see herebelow).

\subsection{Nature of first evolutionary leaps}
We have also suggested applications to biology of the new quantum-like mechanics in scale space \cite{Nottale2004A}.

In the fractal model of the tree of life described hereabove \cite{Chaline1999}, we have voluntarily limited ourselves to an analysis of only the chronology of events (see Fig.~\ref{biomedfig1}), independently of the nature of the major evolutionary leaps. The suggestion of a quantum-type mechanics in scale space and of the new concept of complexergy \cite{Nottale2004A,Nottale2007B}, which is a new conservative quantity appearing from the symmetry of the new scale variables (more precisely, of the fractal dimension become variable and considered as a fifth dimension) allows one to reconsider the question. 

One may indeed suggest that life evolution proceeds in terms of increasing quantized complexergy. This would account for the existence of punctuated evolution \cite{Gould1977}, and for the log-periodic behavior of the leap dates, which can be interpreted in terms of probability density of the events, $P=|\psi|^2 \propto \sin^2[\omega \ln(T-T_c)]$. Moreover, one may contemplate the possibility of an understanding of the nature of the events, even though in a rough way as a first step. 

Indeed, one can expect the first formation of a structure at the fundamental level (lowest complexergy), which is generally characterized by only one length-scale (this is the analog in scale space of the left part of Fig.~\ref{duplication} which concerns position space). Moreover, the most probable value for this scale of formation is predicted to be the `middle' of the scale-space, since the problem is similar to that of a quantum particle in a box, with the logarithms of the minimum scale $\lambda_m$ and maximum scale $\lambda_M$ playing the roles of the walls of the box, so that the fundamental level solution has a peak at a scale $\sqrt{\lambda_m \times \lambda_M}$.

The universal boundary conditions are the Planck-length $l_\Pl$ in the microscopic domain and the cosmic scale $\Lu=\Lambda^{-1/2}$ given by the cosmological constant $\Lambda$ in the macroscopic domain (see Sec.~\ref{cosmocst}). From the predicted and now observed value of the cosmological constant, one finds $\Lu/l_{\Pl}=5.3 \times 10^{60}$, so that the mid scale is at $2.3 \times 10^{30} \, l_{\Pl}\approx 40\; \mu$m. A quite similar result is obtained from the scale boundaries of living systems ($\approx$0.5 Angstr\"oms - 30 m). This scale of $40\; \mu$m is indeed a typical scale of living cells. Moreover, the first `prokaryot' cells appeared about three Gyrs ago had only one hierarchy level (no nucleus).

In this framework, a further increase of complexergy can occur only in a quantized way. The second level describes a system with two levels of organization, in agreement with the second step of evolution leading to eukaryots about 1.7 Gyrs ago (second event in Fig.~\ref{biomedfig1}). One expects (in this very simplified model), that the scale of nuclei be smaller than the scale of prokaryots, itself smaller than the scale of eucaryots: this is indeed what is observed.

The following expected major evolutionary leap is a three organization level system, in agreement with the apparition of multicellular forms (animals, plants and fungi) about 1 Gyr ago (third event in Fig.~\ref{biomedfig1}). It is also predicted that the multicellular stage can be built only from eukaryots, in agreement with what is observed. Namely, the cells of multicellulars do have nuclei; more generally, evolved organisms keep in their internal structure the organization levels of the preceeding stages.

The following major leaps correspond to more complicated structures then more complex functions (supporting structures such as exoskeletons, tetrapody, homeothermy, viviparity), but they are still characterized by fundamental changes in the number of organization levels. Moreover, the first steps in the above model are based on spherical symmetry, but this symmetry is naturaly broken at scales larger than $40\; \mu$m, since this is also the scale beyond which the gravitational force becomes larger than the van der Waals force. One therefore expects the evolutionary leaps that follow the apparition of multicellular systems to lead to more complicated structures, such as those of the Precambrian-Cambrian radiation, than can no longer be described by a single scale variable. 

\subsection{Origin of the genetic code}

We therefore intend, in future works, to extend the model to more general symmetries, boundary conditions and constraints. We also emphasize once again that such an approach does not dismiss the role and the importance of the genetic code in biology. On the contrary, we hope that it may help understanding its origin and its evolution. 

Indeed, we have suggested that the various biological morphologies and functions are solutions of macroscopic Schr\"odinger-type equations, whose solutions are quantized according to integer numbers that represent the various conservative quantities of the system. Among these quantities, one expects to recover the basic physical ones, such as energy, momentum, electric charge, etc... But one may also contemplate the possibility of the existence of prime integrals (conservative quantities) which would be specific of biology (or particularly relevant to biology), among which we have suggested the new concept of complexergy, but also new scale `charges' finding their origin in the internal scale symmetries of the biological systems.
\index{complexergy|)}

The quantization of these various quantities means that any such system would be described by a set of integer numbers, so that one may tentatively suggest that only these numbers, instead of a full continuous and detailed information, would have to be included in the genetic code. In this case the process of genetic code reading, protein synthesis, etc... would be a kind of `analogic solutioner' of Schr\"odinger equation, leading to the final morphologies and functions. Such a view also offers a new line of research toward understanding the apparition of the code, namely, the transformation of what was a purely chemical process into a support of information and of its implementation, thanks to the quantization of the exchanges of energy and other conservative quantities.

We intend to develop this approach in future works, in particular by including the scale relativity tools and methods in a system biology framework allowing multiscale integration \cite{Auffray2007A,Nottale2007B}, in agreement with Noble's `biological relativity' \cite{Noble2006} according to which there is no privileged scale in living systems.

\section{Conclusion}

The theory of scale relativity relies on the postulate that the fundamental laws that govern the various physical, biological and other phenomenons find their origin in first principles. In continuity with previous theories of relativity, it considers that the most fundamental of these principles is the principle of relativity itself. The extraordinary success due to the application of this principle, since now four centuries, to position, orientation, motion (and therefore to gravitation) is well known. 

But, during the last decades, the various sciences have been faced to an ever increasing number of new unsolved problems, of which many are linked to questions of scales. It therefore seemed natural, in order to deal with these problems at a fundamental and first principle level, to extend theories of relativity by including the scale in the very definition of the coordinate system, then to account for these scale transformations in a relativistic way.

We have attempted to give in this article a summarized discussion of the various developments of the theory and of its applications.
The aim of this theory is to describe space-time as a continuous manifold without making the hypothesis of differentiability, and to physically constrain its possible geometry by the principle of relativity, both of motion and of scale. This is effectively made by using the physical principles that directly derive from it, namely, the covariance, equivalence and geodesic principles. These principles lead in their turn to the construction of covariant derivatives, and finally to the writing, in terms of these covariant derivatives, of the motion equations under the form of free-like geodesic equations. Such an attempt is therefore a natural extension of general relativity, since the two-times differentiable continuous manifolds of Einstein's theory, that are constrained by the principle of relativity of motion, are particular sub-cases of the new geometry in construction.

Now, giving up the differentiability hypothesis involves an extremely large number of new possible structures to be investigated and described. In view of the immensity of the task, we have chosen to proceed by steps, using presently known physics as a guide. Such an approach is rendered possible by the result according to which the small scale structure which manifest the nondifferentiability are smoothed out beyond some relative transitions toward the large scales.  One therefore recovers the standard classical differentiable theory as a large scale approximation of this generalized approach. But one also obtains a new geometric theory which allows one to understand quantum mechanics as a manifestation of an underlying nondifferentiable and fractal geometry, and finally to suggest generalizations of it and new domains of application for these generalizations.

Now the difficulty with theories of relativity is that they are meta-theories rather than theories of some particular systems. Hence, after the construction of special relativity of motion at the beginning of the twentieth century, the whole of physics needed to be rendered `relativistic' (from the viewpoint of motion), a task that is not yet fully achieved. 
The same is true as regards the program of constructing a fully scale-relativistic science. Whatever be the already obtained successes, the task remains huge, in particular when one realizes that it is no longer only physics that is concerned, but now many other sciences, in particular biology. Its ability to go beyond the frontiers between sciences may be one of the main interests of the scale relativity theory, opening the hope of a refoundation on mathematical principles and on predictive differential equations of a `philosophy of nature' in which physics would no longer be separated from other sciences.\\

\noindent{\bf Acknowledgements} I gratefully thank the organizers of this Colloquium, C. Vidal and J. Smart, for their kind invitation to contribute and for interesting discussions.

 

\begin{thebibliography}{99}

\bibitem{Abbott1981} Abbott L.F. \&  Wise M.B., 1981, {\it Am. J. Phys.} {\bf 49}, 37.
%
\bibitem{Agnese1997} Agnese A.G., Festa R., 1997, \textit{Phys. Lett.} \textbf{A 227}, 165.
%
\bibitem{Allegre1982} All\`egre C., Le Mouel J. \& Provost A., 1982, \textit{Nature} \textbf{297}, 47
%
\bibitem{Amelino2001} Amelino-Camelia G., 2001, {Phys. Lett.} \textbf{B 510}, 255.
%
\bibitem{Amelino2002} Amelino-Camelia G., 2002, {Int. J. Mod. Phys.} \textbf{D 11}, 1643.
%
\bibitem{Auffray2007A} Auffray Ch. \& Nottale L., 2008, {\it Progr. Biophys. Mol. Bio.},  \textbf{97}, 79.%
%
\bibitem{BenAdda2000} Ben Adda F. \& Cresson J., 2000, {\it C. R. Acad. Sci. Paris} {\bf 330}, 261.
%
\bibitem{BenAdda2004} Ben Adda F. \& Cresson J., 2004, {\it Chaos, Solitons \& Fractals} {\bf 19}, 1323.
%
\bibitem{BenAdda2005} Ben Adda F. \& Cresson J., 2005, {\it Applied Mathematics and Computation} {\bf 161}, 323.
%
\bibitem{Berry1996} Berry M.V., 1996, {\it J. Phys. A: Math. Gen.} {\bf 29}, 6617.%
%
\bibitem{Cafiero1995} Cafiero, R., Loreto, V., Pietronero, L., Vespignani, A., Zapperi, S., 1995, {\it Europhys. Lett.} {\bf 29}, 111.
%
\bibitem{Campagne2003} Campagne J.E. \& Nottale L., 2003, unpublished preprint.
%
\bibitem{Carpinteri1996} Carpinteri, A., Chiaia, B., 1996, {\it Chaos, Solitons \& Fractals} {\bf 7}, 1343.
%
\bibitem{Cash2002} Cash R., Chaline J., Nottale L., Grou P., 2002, \textit{C.R. Biologies}  \textbf{325}, 585.
%
\bibitem{Castro1997} Castro C., 1997, {\it Found. Phys. Lett.} {\bf 10}, 273.
%
\bibitem{Castro2000} Castro C. \& Granik A., 2000, {\it Chaos, Solitons \& Fractals} {\bf 11}, 2167.
%
\bibitem{Celerier2004} C\'el\'erier M.N. \& Nottale L.,  2004,  {\it J. Phys. A: Math. Gen.} {\bf 37}, 931.
%
\bibitem{Celerier2006} C\'el\'erier M.N. \& Nottale L.,  2006,  {\it J. Phys. A: Math. Gen.} {\bf 39}, 12565.%
%
\bibitem{Chaline1999} Chaline J., Nottale L. \& Grou P., 1999, \textit{C.R. Acad. Sci. Paris}, \textbf{328}, 717.
%
\bibitem{Connes1994} Connes A., 1994, \textit{Noncommutative Geometry} (Academic Press, New York).
%
\bibitem{Connes1998} Connes A., Douglas M.R. \& Schwarz A., {\it J. High Energy Phys.} {\bf 02}, 003 (hep-th/9711162).
%
\bibitem{Cresson2001} Cresson J., 2001, \textit{M\'emoire d'habilitation \`a diriger des recherches}, Universit\'e de Franche-Comt\'e, Besan{\c c}on.
%
\bibitem{Cresson2002} Cresson J., 2002, {\it Chaos, Solitons \& Fractals} {\bf 14}, 553.
%
\bibitem{Cresson2003} Cresson, J., 2003, {\it J. Math. Phys.} {\bf 44}, 4907.
%
\bibitem{Cresson2006} Cresson, J., 2006, {\it International Journal of Geometric Methods in Modern Physics}, {\bf 3}, no. 7.
%
\bibitem{Cresson2007} Cresson, J., 2007, {\it J. Math. Phys.} {\bf 48}, 033504.
%
\bibitem{daRocha2003} da Rocha D. \& Nottale L., 2003, \textit{Chaos, Solitons \& Fractals} \textbf{16}, 565.
%
\bibitem{Dubois2000} Dubois D., 2000, in {\it Proceedings of CASYS'1999}, 3rd International Conference on Computing Anticipatory Systems, Li\`ege, Belgium, \textit{Am. Institute of Physics Conference Proceedings} {\bf 517}, 417.
%
\bibitem{Dubrulle1997} Dubrulle B., Graner F. \& Sornette D. (Eds.), 1997,  in {\it Scale invariance and beyond}, Proceedings of Les Houches school, B. Dubrulle, F. Graner \& D. Sornette eds., (EDP Sciences, Les Ullis/Springer-Verlag, Berlin, New York), p. 275.

\bibitem{ElNaschie1992} El Naschie M.S., 1992, {\it Chaos, Solitons \& Fractals} {\bf 2}, 211.
%
\bibitem{ElNaschie1995} El Naschie M.S., R\"ossler 0. \& Prigogine I. (Eds.), 1995,  \textit{Quantum Mechanics, Diffusion and Chaotic Fractals}, Pergamon, New York.%
%
\bibitem{ElNaschie2000} El Naschie M.S.,  {\it Chaos, Solitons \& Fractals} {\bf 11}, 2391.
%
\bibitem{Feynman1965} Feynman R.P. \& Hibbs A.R., 1965, {\it Quantum Mechanics and Path Integrals} (MacGraw-Hill, New York).
%
\bibitem{Forriez2005}  Forriez M., 2005, {\it Etude de la Motte de Boves}, Geography and History Master I report, Artois University.
%
\bibitem{Forriez2006}  Forriez M. \& Martin P., 2006, {\it Geopoint Colloquium: Demain la G\'eographie},  in press.
%
\bibitem{Forriez2007}  Forriez M., 2005, {\it Construction d'un espace g\'eographique fractal}, Geography Master II report, Avignon University.
%
\bibitem{Galopeau2004} Galopeau P., Nottale L., Ceccolini D., Da Rocha D., Schumacher G. \& Tran-Minh N., 2004, in \textit{Scientific Highlights 2004, Proceedings of the Journ\'ees de la SF2A}, Paris 14-18 June 2004, F. Combes, D. Barret, T. Contini, F. Meynadier \& L. Pagani (eds.), EDP Sciences, p. 75.
%
\bibitem{Georgi1974A} Georgi H. \& Glashow S.L., 1974, {\it Phys Rev. Lett.} {\bf 32}, 438.
%
\bibitem{Georgi1974B} Georgi H., Quinn H.R. \& Weinberg S., 1974, {\it Phys Rev. Lett.} {\bf 33}, 451.
%
\bibitem{Glashow1961} Glashow S.L., 1961, {\it Nucl. Phys.} {\bf 22}, 579.
%
\bibitem{Gould1977} Gould S.J. \& Eldredge N., 1977, \textit{Paleobiology}, \textbf{3}(2), 115.
%
\bibitem{Grabert1979} Grabert H., H\"anggi P. \& Talkner P., 1979, {\it Phys. Rev.} {\bf A19}, 2440.
%
\bibitem{Green1987} Green M. B., Schwarz J. H. \& Witten E., 1987, {\it Superstring Theory} (2 vol.), Cambridge University Press.
%
\bibitem{Grou1987} Grou P., 1987, {\it L'aventure \'economique}, L'Harmattan, Paris.
%
\bibitem{Grou2004} Grou P., Nottale L. \& Chaline J., 2004, in {\it Zona Arqueologica, Miscelanea en homenaje a Emiliano Aguirre}, IV Arqueologia, 230, Museo Arquelogico Regional, Madrid.
%
\bibitem{Hall2004} Hall M.J.W., 2004, {\it J. Phys. A: Math. Gen.} {\bf 37}, 9549.%
%
\bibitem{Hermann1997} Hermann R., 1997, \textit{J. Phys. A: Math. Gen.} \textbf{30}, 3967.
%
\bibitem{Hermann1998} Hermann R., Schumacher G. \& Guyard R., 1998, \textit{Astron. Astrophys.} {\bf 335}, 281.
%
\bibitem{Hinshaw2008} Hinshaw G. {\it et al.}, Five-Years WMAP observations, arXiv:08030732 [astro-ph]%
%
\bibitem{Johansen2001} Johansen A. \& Sornette D., 2001, \textit{Physica} \textbf{A 294}, 465.
%
\bibitem{Jumarie2001} Jumarie G., 2001, {\it Int. J. Mod. Phys. } {\bf A 16}, 5061.%
%
\bibitem{Jumarie2006} Jumarie G., 2006, {\it Computer and Mathematics} {\bf 51}, 1367.
%
\bibitem{Jumarie2006A} Jumarie G., 2006, {\it Chaos, Solitons \& Fractals} {\bf 28}, 1285.
%
\bibitem{Jumarie2007} Jumarie G., 2007, {\it Phys. Lett.} {\bf A}, {bf 363}, 5.
%
\bibitem{Kroger2000} Kr\"oger H., 2000, {\it Phys. Rep.} {\bf 323}, 81.
%
\bibitem{Lang1980} Lang K.R., 1980, {\it Astrophysical Formulae}, Springer-Verlag.
%
\bibitem{Laperashvili2001} Laperashvili L.V. \& Ryzhikh D.A., 2001, arXiv: hep-th/0110127 (Institute for Theoretical and Experimental Physics, Moscow).
%
\bibitem{LevyLeblond1976} Levy-Leblond J.M., 1976, \textit{Am. J. Phys.} \textbf{44}, 271.
%
\bibitem{Losa2002} Losa G., Merlini D., Nonnenmacher T. \& Weibel E. (Editors) \textit{Fractals in Biology and Medicine, Vol. III}, Proceedings of Fractal 2000 Third International Symposium, Birkh\"auser Verlag.
%
\bibitem{Mandelbrot1982} Mandelbrot B., 1982, {\it The Fractal Geometry of Nature}, Freeman, San Francisco.
%
\bibitem{Martin2006}  Martin P. \& Forriez M., 2006, {\it Geopoint Colloquium: Demain la G\'eographie},  in press.
%
\bibitem{McKeon1992} McKeon D.G.C. \& Ord G. N., 1992, {\it Phys. Rev. Lett.} {\bf 69}, 3.
%
\bibitem{Nelson1966} Nelson E., 1966, {\it Phys. Rev.} {\bf 150}, 1079.
%
\bibitem{Noble2002} Noble D., 2002, {\it BioEssays}, {\bf 24}, 1155.
%
\bibitem{Noble2006} Noble D., 2006, {\it The music of Life: Biology Beyond the Genome}, Oxford University Press.
%
\bibitem{Nottale1984} Nottale L. \& Schneider J., 1984, {\it J. Math. Phys.} {\bf 25}, 1296.
%
\bibitem{Nottale1989} Nottale L., 1989, {\it Int. J. Mod. Phys. A} {\bf 4}, 5047.
%
\bibitem{Nottale1992} Nottale L., 1992, {\it Int. J. Mod. Phys. A} {\bf 7}, 4899.
%
\bibitem{Nottale1993} Nottale L., 1993, {\it Fractal Space-Time and Microphysics: Towards a Theory of Scale Relativity}, World Scientific, Singapore.
%
\bibitem{Nottale1994B} Nottale L., 1994, in {\it Relativity in General}, (Spanish Relativity Meeting 1993), edited J. Diaz Alonso \& M. Lorente Paramo (Eds.), Editions Fronti\`eres, Paris, p. 121.
%
\bibitem{Nottale1994C} Nottale L., 1994, in \textit{Chaos and Diffusion in Hamiltonian Systems}, Proceedings of the fourth workshop in Astronomy and Astrophysics of Chamonix (France), 7-12 February 1994, D. Benest \& C. Froeschl\'e (Eds.), Editions Fronti\`eres, p. 173.
%
\bibitem{Nottale1996A} Nottale L., 1996, {\it Chaos, Solitons \& Fractals} {\bf 7}, 877.
%
\bibitem{Nottale1996B} Nottale L., 1996, \textit{Astron. Astrophys. Lett.} \textbf{315}, L9.
%
\bibitem{Nottale1997A} Nottale L., 1997, {\it Astron. Astrophys.} {\bf 327}, 867.
%
\bibitem{Nottale1997B} Nottale L., 1997, in {\it Scale invariance and beyond}, Proceedings of Les Houches school, B. Dubrulle, F. Graner \& D. Sornette eds., (EDP Sciences, Les Ullis/Springer-Verlag, Berlin, New York), p. 249.
%
\bibitem{Nottale1997C} Nottale L., Schumacher G. \& Gay J., 1997, \textit{Astron. Astrophys.} \textbf{322}, 1018.
%
\bibitem{Nottale1998A} Nottale L., 1998, \textit{Chaos, Solitons \& Fractals} {\bf 9}, 1035.
%
\bibitem{Nottale1998B} Nottale L., 1998, \textit{Chaos, Solitons \& Fractals} {\bf 9}, 1043.
%
\bibitem{Nottale1998E} Nottale L. \& Schumacher G., 1998, in {\it Fractals and beyond: complexities in the sciences}, M. M. Novak (Ed.), World Scientific, p. 149.
%
\bibitem{Nottale1999} Nottale L., 1999, \textit{Chaos, Solitons \& Fractals} \textbf{10}, 459.
%
\bibitem{Nottale2000C} Nottale L., Schumacher G. \& Lef\`evre E.T., 2000, \textit{Astron. Astrophys.} \textbf{361}, 379.
%
\bibitem{Nottale2001A} Nottale L., 2001, in {\it Frontiers of Fundamental Physics}, Proceedings of Birla Science Center Fourth International Symposium, 11-13 December 2000, B. G. Sidharth \& M. V. Altaisky (Eds.), Kluwer Academic, p. 65.
%
\bibitem{Nottale2001B} Nottale L., 2001, \textit{Revue de Synth\`ese}, \textbf{T. 122}, 4e S., No 1, January-March 2001, p. 93.
%
\bibitem{Nottale2001C} Nottale L., 2001, \textit{Chaos, Solitons \& Fractals} \textbf{12}, 1577.
%
\bibitem{Nottale2000B} Nottale L., Chaline J. \& Grou P., 2000, \textit{Les arbres de l'\'evolution: Univers, Vie, Soci\'et\'es}, Hachette, Paris, 379 pp.
%
\bibitem{Nottale2002B} Nottale L., Chaline J. \& Grou P., 2002, in \textit{Fractals in Biology and Medicine, Vol. III}, Proceedings of Fractal 2000 Third International Symposium, G. Losa, D. Merlini, T. Nonnenmacher \& E. Weibel (Eds.), Birkh\"auser Verlag, p. 247.
%
\bibitem{Nottale2002D} Nottale L., 2002, in {\it Penser les limites. Ecrits en l'honneur d'Andr\'e Green}, Delachiaux et Niestl\'e, p. 157.
%
\bibitem{Nottale2003A} Nottale L., 2003, \textit{Chaos, Solitons \& Fractals} \textbf{16}, 539.
%
\bibitem{Nottale2004A} Nottale L., 2004, {\it American Institute of Physics Conference Proceedings} {\bf 718}, 68.%
%
\bibitem{Nottale2004B} Nottale L., 2004, in ``Computing Anticipatory Systems. CASYS'03 - Sixth International Conference" (Li\`ege, Belgique, 11-16 August 2003), Daniel M. Dubois Editor, {\it American Institute of Physics Conference Proceedings} {\bf 718}, 68.%
%
\bibitem{Nottale2004BB} Nottale L., 2004, in \textit{Scientific Highlights 2004, Proceedings of the Journ\'ees de la SF2A}, Paris 14-18 June 2004, F. Combes, D. Barret, T. Contini, F. Meynadier \& L. Pagani (Eds.), EDP Sciences, p. 699.
%
\bibitem{Nottale2006} Nottale L., C\'el\'erier M.N. \& Lehner T., 2006, \textit{J. Math. Phys.} \textbf{47}, 032303.
%
\bibitem{Nottale2006B} Nottale L., 2006, in Proceedings of the Sixth International Symposium {\it Frontiers of Fundamental Physics}, Udine, Italy, 26-29 September 2004, Sidharth, B.G., Honsell, F. \& de Angelis, A. (Eds.), Springer, p. 107.
%
\bibitem{Nottale2006A} Nottale L., 2008, in ``Proceedings of 7th International Colloquium on Clifford Algebra", Toulouse, France, 19-29 Mai 2005, {\it Advances in Applied Clifford Algebra}, {\bf 18}, 917.
%
\bibitem{Nottale2006D} Nottale L. \& Timar P., 2006, {\it Psychanalyse et Psychose} {\bf 6}, 195.
%
\bibitem{Nottale2007A} Nottale L. and C\'el\'erier M.N., 2007, {\it J. Phys. A}, {\bf40}, 14471.%
%
\bibitem{Nottale2007G} Nottale L., 2007, Special Issue (July 2007) on ``Physics of Emergence and Organization", {\it Electronic Journal of Theoretical Physics} {\bf 4} No. 15, in press.
%
\bibitem{Nottale2007H} Nottale L., H\'eliodore F. \& Dubois J., 2007, {\it Log-periodic evolution of earthquake rate}, in preparation.
%
\bibitem{Nottale2007I} Nottale L., 2007, in {\it Les grands d\'efis technologiques et scientifiques du XXI\`e si\`ecle}, P. Bourgeois et P. Grou Eds., Ellipse, Chap. 9, p. 121.
%
\bibitem{Nottale2007L} Nottale L., 2007, in {\it Les Grands D\'efis Technologiques et Scientifiques au XXI\`e si\`ecle}, under the direction of Philippe Bourgeois and Pierre Grou, Ellipses, Chap. 9, p. 121.%
%
\bibitem{Nottale2007B} Nottale L. \& Auffray C., 2008, {\it Progr. Biophys. Mol. Bio.}, {\bf 97}, 115.
%
\bibitem{Nottale2008A} Nottale L. and Timar P., 2008, in {\it Simultaneity: Temporal Structures and Observer Perspectives}, Susie Vrobel, Otto E. Rössler, Terry Marks-Tarlow, Eds., World Scientific, Singapore), Chap. 14, p. 229.
%
\bibitem{Nottale2007} Nottale L., 2008, {\it The theory of Scale Relativity}, book 540pp, submitted for publication.
%
\bibitem{Novak1998} Novak M. (Editor), 1998, {\it Fractals and Beyond: Complexities in the Sciences, Proceedings of the Fractal 98 conference}, World Scientific.
%
\bibitem{Ord1983} Ord G.N., 1983, {\it J. Phys. A: Math. Gen.} {\bf 16}, 1869.
%
\bibitem{Ord1996} Ord G.N., 1996, {\it Ann. Phys.} {\bf 250}, 51.
%
\bibitem{Ord2002} Ord G.N. \& Galtieri J.A., 2002, \textit{Phys. Rev. Lett.} \textbf{1989}, 250403.
%
\bibitem{PDG2006} (Particle Data Group) Yao W.-M. et al., 2006, {\it J. Phys.} {\bf G 33}, 1
%
\bibitem{Pecker1959} Pecker J.C. \& Schatzman E., 1959, {\it Astrophysique g\'en\'erale}, Masson, Paris.
%
\bibitem{Peterson1979} Peterson M.A., 1979, \textit{Am. J. Phys.} \textbf{47}, 488.
%
\bibitem{Pissondes1999} Pissondes J.C., 1999, \textit{J. Phys. A: Math. Gen.} \textbf{32}, 2871.
%
\bibitem{Polchinski1998} Polchinski J, 1998, {\it String Theories}, Cambridge University Press, Cambridge UK.
%
\bibitem{Queiros2000} Queiros-Cond\'e D., 2000, {\it C. R. Acad. Sci. Paris} {\bf 330}, 445.
%
\bibitem{Rovelli1988} Rovelli C. \& Smolin L., 1988, \textit{Phys.Rev. Lett.} \textbf{61}, 1155.
%
\bibitem{Rovelli1995} Rovelli C. \& Smolin L., 1995, \textit{Phys.Rev.} \textbf{D 52}, 5743.
%
\bibitem{Saar1999} Saar S.H. \& Brandenburg A., 1999, {\it Astrophys. J.} {\bf 524}, 295.
%
\bibitem{Salam1968} Salam A., 1968, {\it Elementary Particle Theory}, Eds. N. Svartholm, Almquist, \& Wiksells, Stockholm.
%
\bibitem{Schneider2008} Schneider J., 2008, {\it The Extrasolar Planets Encyclopaedia}, http://exoplanet.eu/index.php%
%
\bibitem{Smith2004} Smith N.P, Nickerson D.P., Crampin E.J. \& Hunter P.J., 2004, {\it Acta Numerica}, p. 371, Cambridge University Press.
%
\bibitem{Sornette1995} Sornette D. \& Sammis C.G., 1995, {\it  J. Phys. I France} {\bf 5}, 607.
%
\bibitem{Sornette1996} Sornette D., Johansen A. \& Bouchaud J.P., 1996, {\it J. Phys. I France} {\bf 6}, 167.
%
\bibitem{Sornette1998} Sornette D., 1998, \textit{Phys. Rep.} \textbf{297}, 239.
%
\bibitem{Spergel2006} D.N. Spergel et al., 2006, arXiv: astro-ph/0603449
%
\bibitem{Tegmark2006} Tegmark M. et al., 2006, \textit{Phys. Rev.} \textbf{D 74}, 123507 (arXiv: astro-ph/0308632).
%
\bibitem{Timar2002} Timar P., 2002, http://www.spp.asso.fr/main/PsychanalyseCulture/
SciencesDeLaComplexite/Items/3.htm.
%
\bibitem{Wang1993} Wang M.S. \& Liang W.K., 1993, {\it Phys. Rev.} {\bf D 48}, 1875.
%
\bibitem{Weinberg1967} Weinberg S., 1967, {\it Phys. Rev. Lett.} {\bf 19}, 1264.
%
\bibitem {Zeldovich1983} Zeldovich Ya. B., Ruzmaikin A.A. \& Sokoloff D.D., 1983, {\it Magnetic Fields in Astrophysics}, Gordon and Breach.
%
%
\end{thebibliography}
\end{document}